\documentclass{aa} 
\usepackage{graphicx} 
\usepackage[varg]{txfonts}
\usepackage{natbib}
\bibpunct{(}{)}{;}{a}{}{,}

\newcommand{\code}[1]{\texttt{#1}}

\def\farcs{\hbox{$.\!\!^{\prime\prime}$}} 
\newcommand{\HII}{H\,{\small II}}
\newcommand{\Ha}{H{$\alpha$}}
\newcommand{\TpeakCO}{$T_{\mathrm{peak}}^{\mathrm{CO}\:3-2}$}
\newcommand{\WHCOp}{$W_{4-3}^{\mathrm{HCO}^+}$}

\makeatletter
\renewcommand*\aa@pageof{, page \thepage{} of \pageref*{LastPage}}
\makeatother

\begin{document}

\title{The extremely sharp transition between molecular and ionized gas in the Horsehead nebula}
\author{C. Hernández-Vera\inst{\ref{inst1}}\orcid{0009-0009-2320-7243}
\and V. V. Guzmán\inst{\ref{inst1},\ref{inst2}}\orcid{0000-0003-4784-3040}
\and J. R. Goicoechea\inst{\ref{inst3}}\orcid{0000-0001-7046-4319}
\and V. Maillard\inst{\ref{inst4}}
\and J. Pety\inst{\ref{inst5},\ref{inst6}}
\and F. Le Petit\inst{\ref{inst4}}
\and M. Gerin\inst{\ref{inst6}}
\and E.~Bron\inst{\ref{inst4}}
\and E. Roueff\inst{\ref{inst4}}
\and A. Abergel\inst{\ref{inst7}}
\and T. Schirmer\inst{\ref{inst8}}
\and J. Carpenter\inst{\ref{inst9}}
\and P. Gratier\inst{\ref{inst10}}
\and K. Gordon\inst{\ref{inst11}}
\and K. Misselt\inst{\ref{inst12}}}

\institute{Instituto de Astrof\'isica, Pontificia Universidad Cat\'olica de Chile, Av. Vicu\~na Mackenna 4860, 7820436 Macul, Santiago, Chile \email{chernandez@astro.puc.cl}\label{inst1}
\and N\'ucleo Milenio de Formaci\'on Planetaria (NPF), Chile\label{inst2}
\and Instituto de Física Fundamental (CSIC), Calle Serrano 121, 28006, Madrid, Spain\label{inst3}
\and LERMA, Observatoire de Paris, PSL Research University, CNRS, Sorbonne Université, 92190 Meudon, France\label{inst4}
\and IRAM, 300 rue de la Piscine, 38406 Saint-Martin-d’Hères, France\label{inst5}
\and LERMA, Observatoire de Paris, PSL Research University, CNRS, Sorbonne Université, 75014 Paris, France\label{inst6}
\and Institut d’Astrophysique Spatiale, Université Paris-Saclay, CNRS, Bâtiment 121, 91405 Orsay Cedex, France\label{inst7}
\and Department of Space, Earth and Environment, Chalmers University of Technology, Onsala Space Observatory, 439 92 Onsala, Sweden\label{inst8}
\and Joint ALMA Observatory, Alonso de Cordova 3107 Vitacura, Santiago, Chile\label{inst9}
\and Laboratoire d’Astrophysique de Bordeaux, Univ. Bordeaux, CNRS, B18N, Allée Geoffroy Saint-Hilaire, 33615 Pessac, France\label{inst10}
\and Space Telescope Science Institute, 3700 San Martin Drive, Baltimore, MD 21218-2463, USA\label{inst11}
\and Steward Observatory, University of Arizona, Tucson, AZ, 85721, USA\label{inst12}
} 

\date{Received 16 June 2023 /
Accepted 10 July 2023 }

\abstract{
Massive stars can determine the evolution of molecular clouds by eroding and photo-evaporating their surfaces with strong ultraviolet (UV) radiation fields. Moreover, UV radiation is relevant in setting the thermal gas pressure in star-forming clouds, whose influence can extend across various spatial scales, from the rims of molecular clouds to entire star-forming galaxies. Probing the fundamental structure of nearby molecular clouds is therefore crucial to understand how massive stars shape their surrounding medium and how fast molecular clouds are destroyed, specifically at their UV-illuminated edges, where models predict an intermediate zone of neutral atomic gas between the molecular cloud and the surrounding ionized gas whose size is directly related to the exposed physical conditions. We present the highest angular resolution ($\sim$0\farcs5, corresponding to $207$~au) and velocity-resolved images of the molecular gas emission in the Horsehead nebula, using CO $J=3-2$ and HCO$^+$ $J=4-3$ observations with the Atacama Large Millimeter/submillimeter Array (ALMA). We find that CO and HCO$^+$ are present at the edge of the cloud, very close to the ionization (H$^+$/H) and dissociation fronts (H/H$_2$), suggesting a very thin layer of neutral atomic gas ($<650$~au) and a small amount of CO-dark gas ($A_{\mathrm{V}}=0.006-0.26$ mag) for stellar UV illumination conditions typical of molecular clouds in the Milky Way. The new ALMA observations reveal a web of molecular gas filaments with an estimated thermal gas pressure of $P_{\mathrm{th}} = (2.3 - 4.0) \times 10^6$~K~cm$^{-3}$, and the presence of a steep density gradient at the cloud edge that can be well explained by stationary isobaric photo-dissociation region (PDR) models with pressures consistent with our estimations. However, in the \HII{} region and PDR interface, we find $P_{\mathrm{th,PDR}} > P_{\mathrm{th,\HII{}}}$, suggesting the gas is slightly compressed. Therefore, dynamical effects cannot be completely ruled out and even higher angular observations will be needed to unveil their role.}


\maketitle 

\section{Introduction}

\begin{figure*}[t!]
\centering
\includegraphics[width=0.9\linewidth]{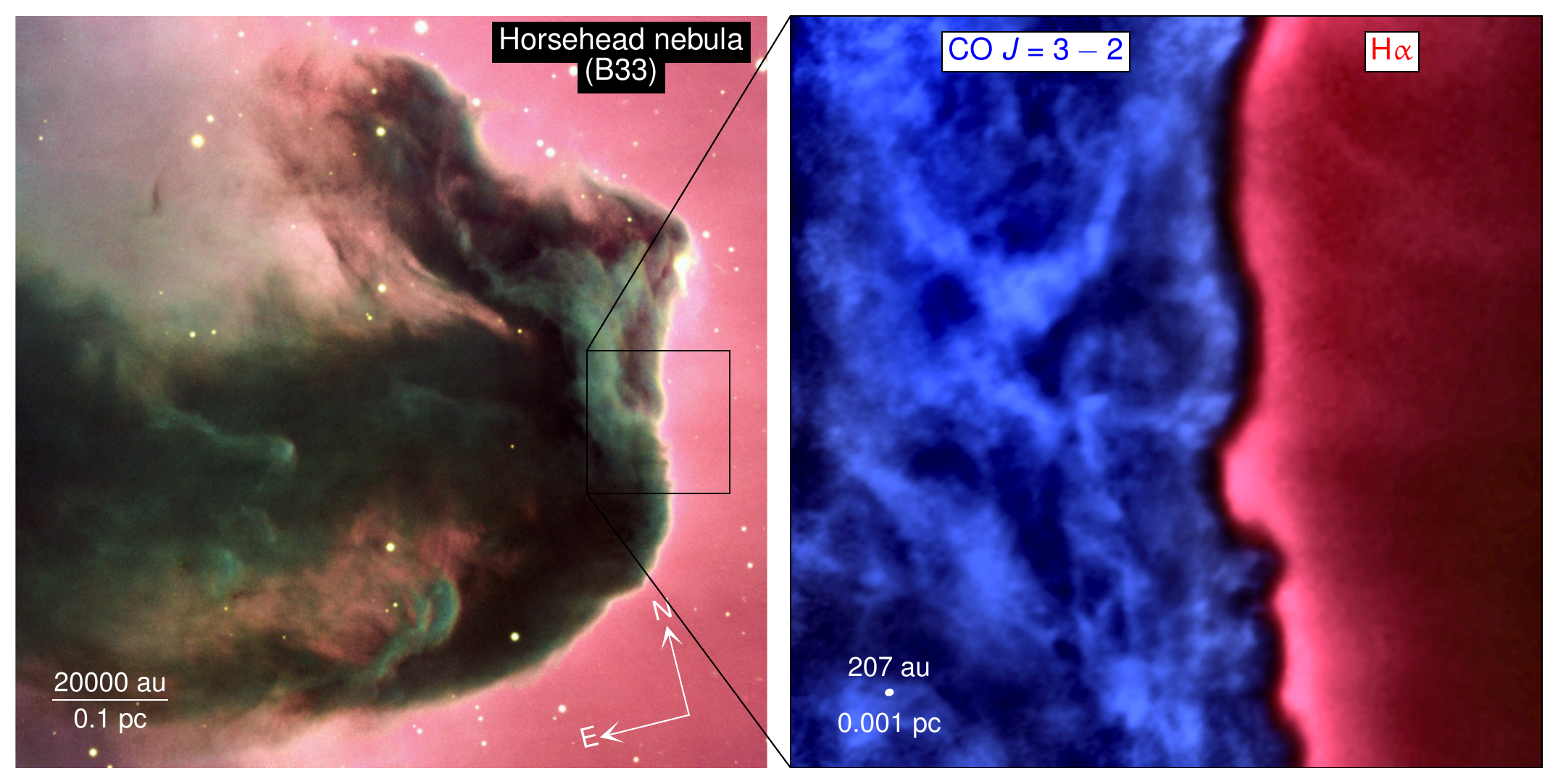}
\caption{Multiphase view of the Horsehead nebula (Barnard 33). Left: Composite color image of the Horsehead nebula and part of the \HII{} region IC 434 observed with VLT (ESO). The size of the image is about $7^\prime \times 7^\prime$, where $1^{\prime}$ is equal to $0.12$~pc at the Horsehead distance. Right: Zoom of the edge of the molecular cloud imaged with ALMA in the CO $J=3-2$ line emission (blue). The hot ionized gas coming from the \HII{} region is traced by the \Ha{} line emission imaged with the $0.9$m KPNO telescope \citep{Pound2003}, resolving scales of about $1^{\prime\prime}$ (red). The field of view in this case is approximately $50^{\prime\prime} \times 50^{\prime\prime}$ or, equivalently, $0.1$~pc $\times\:0.1$~pc. The synthesized beam size of the ALMA observations is shown in the bottom left corner, as well as the corresponding physical scale. The dark region between the right edge of the CO emission and the bright rim of \Ha{} is the neutral atomic layer.}
\label{fig:hh-multiphase}
\end{figure*}

Radiative and mechanical interactions between massive stars and molecular clouds are fundamental to understand the evolution of galaxies. As a consequence of their strong radiation fields and winds, massive stars are able to trigger \citep{Elmegreen1997,Luisi2021} or quench \citep{Walch2012,Kim2018,Pabst2019} the formation of new stars by compressing or disrupting their natal clouds, respectively. It is therefore of paramount importance to study the structure of molecular clouds exposed to stellar ultraviolet (UV) radiation fields, whose interaction occurs in so-called photo-dissociation regions \citep[PDRs,][]{Hollenbach1999}, to assess the effects of stellar feedback. In general, molecular clouds are far from homogeneous sources; they are rather complex structures that include filaments, shells, and pillars \citep{Hacar2022,Pineda2022}, but their fundamental structure at sub-parsec scales ($<0.1$~pc) has been poorly constrained in the context of star formation, specifically in the feedback-impacted surface of molecular clouds in star-forming regions.

Furthermore, a chemical layering is expected at the UV-illuminated surfaces of molecular clouds. The extreme-UV (EUV, $h \nu >13.6$ eV) photons from massive stars produce an adjacent \HII{} region where the gas is fully ionized, while far-UV (FUV, $6 < h \nu < 13.6$ eV) photons interact with neutral gas in the outer layers of molecular clouds, photo-dissociating molecules and thus producing a PDR \citep{Hollenbach1999,Wolfire2022}. The boundary between the \HII{} region and the PDR is delimited by the ionization front, corresponding to the transition from ionized to neutral atomic hydrogen (H$^+$/H). Deeper into the molecular cloud, FUV photons are attenuated by gas and dust \citep{Sternberg2014}. As a result, PDRs are characterized by the transitions from atomic to molecular gas (H/H$_2$ and C$^+$/C/CO transition zones). The relative locations of these transition zones, commonly expressed in terms of visual extinction $A_{\mathrm{V}}$, depend on the ratio between the incident FUV radiation field ($G_0$) and the hydrogen nucleus number density ($n_{\mathrm{H}}$), as well as on the metallicity and dust grain properties \citep{Goicoechea2022b,Wolfire2022}.

Attempts to reproduce the structure of the UV-illuminated surface layer of molecular clouds and to locate the transitions between ionized, atomic, and molecular gas have been made with stationary models using, in general, a constant gas density or constant pressure \citep{Tielens1985,Sternberg1989,Kaufman1999,LePetit2006}. However, the constant gas density assumption often predicts larger separations between the H/H$_2$ and C$^+$/C/CO transition zones than what has been measured empirically in recent years \citep{Goicoechea2016}. Instead, models with constant pressure are, in general, more consistent with the observations but with relatively high thermal gas pressures that cannot be solely explained by the pressure equilibrium with the adjacent \HII{} region, and thus dynamical effects such as photoevaporation are often invoked \citep{Bron2018,Joblin2018}. This has strong implications for the amount of "CO-dark H$_2$ gas" predicted by models, which corresponds to the fraction of molecular gas that is not fully traced by CO emission \citep{Grenier2005,Wolfire2010}, and thus for predictions of the CO-to-H$_2$ conversion factor. Therefore, performing detailed observations of molecular clouds, where we can resolve the different transition zones, is of great interest to have a good estimate of the amount of material available to form new stars, which can be linked to the star formation rate (SFR) by the Kennicutt-Schmidt law \citep{Schmidt1959,Kennicutt1998} as well as to the evolution of galaxies \citep{Tacconi2020}.

\begin{table*}
\caption{Observational parameters for the maps shown in Figs.~\ref{fig:hh-grid} and \ref{fig:hh-channelmaps}.}              
\label{table:1}
\centering
\begin{tabular}{c c c c c c}          
\hline\hline
Line & Frequency & Chan. Width & Beam & PA & Chan. rms\\ 
& (GHz) & (km~s$^{-1}$) & ($^{\prime\prime}$) & $^\circ$ & (mJy~beam$^{-1}$)\\
\hline                                   
CO $J=3-2$ & $345.796$ & $0.22$ & $0.68$ $\times$ $0.53$ & $-74.41$ & $5.28$\\
HCO$^+$ $J=4-3$ & $356.734$ & $0.22$ & $0.66$ $\times$ $0.52$ & $-75.83$ & $5.43$\\
\hline
\end{tabular}
\end{table*}

\begin{figure*}[ht]
\centering
\includegraphics[width=0.92\linewidth]{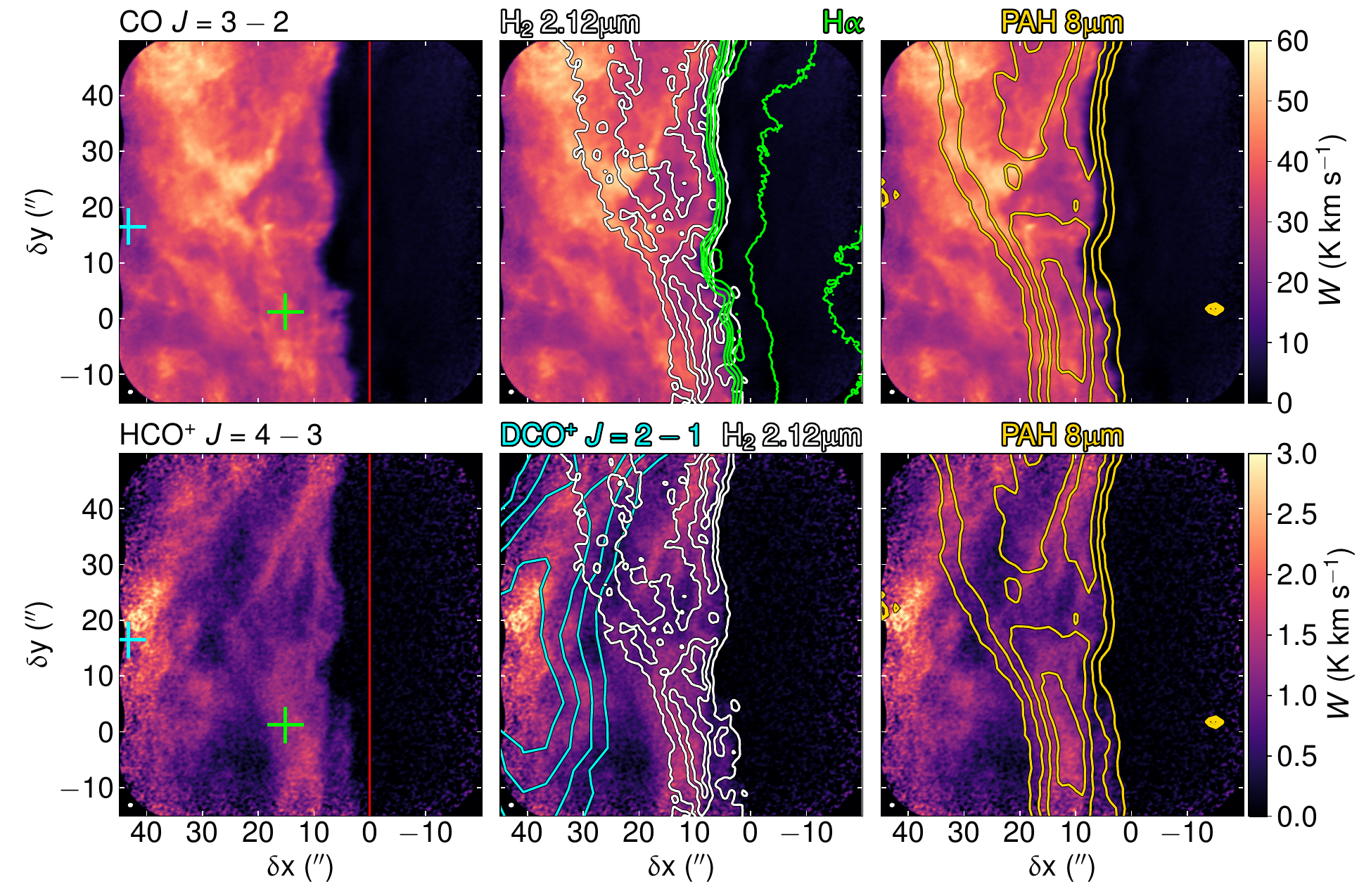}
\caption{ALMA high angular resolution images of the Horsehead nebula. Top: Integrated intensity map of the CO $J=3-2$ line. Bottom: Integrated intensity map of the HCO$^+$ $J=4-3$ line. The green contours represent the \Ha{} line emission \citep{Pound2003}. The cyan contours show the DCO$^{+}$ $J=2-1$ line emission \citep{Pety2007}. The white contours represent the H$_2$ 2.12 $\mu$m rovibrational line emission \citep{Habart2005} and the yellow contours represent the PAH $8$ $\mu$m emission \citep{Schirmer2020}. The horizontal zero (red vertical line) delineates the PDR edge. The cyan and green crosses show the dense core \citep{Pety2007} and PDR \citep{Gerin2009} positions, respectively. Maps have been rotated $14^\circ$ counterclockwise to bring the illuminating star direction in the horizontal direction. The angular resolution of ALMA observations is shown in the bottom left corner.}
\label{fig:hh-grid}
\end{figure*}

Our most detailed images of the edge of a molecular cloud to date are the ones of the famous Orion Bar PDR \citep{Goicoechea2016,Habart2023}. Submillimeter observations from the Atacama Large Millimeter/submillimeter Array (ALMA) together with near-infrared (NIR) images using adaptive optics with Keck telescope, and recent James Webb Space Telescope (JWST) observations using the NIRCam instrument as part of an Early Release Science program \citep[PDRs4All,][]{Berne2022}, revealed high-density filamentary substructures, suggesting that dynamical effects could be important at the UV-illuminated molecular cloud surface. Moreover, the observed H/H$_2$ and C$^+$/C/CO transitions were found to be very close and were not spatially resolved, which could be related to dynamical effects \citep{Goicoechea2016}. The Orion Bar is, however, illuminated by an extremely high-UV radiation field \citep{Marconi1998} \citep[$G_0 \sim 10^4$, where $G_0$ equal to $1.7$ is the average radiation field in the local interstellar medium;][] {Draine1978}. Such a high radiation field is not representative of most of the UV-illuminated molecular gas in both the Milky Way and normal galaxies, which are mostly illuminated by a low to moderate radiation field. Consequently, probing the structure of gas interacting with a moderate radiation field is crucial to get a broader view of the role of photo-ionization and winds caused by massive stars in the structure and evolution of molecular clouds.

Due to its proximity \citep[$\sim 400$~pc, ][]{Anthony-Twarog1982} and favored geometry, the PDR located at the edge of the Horsehead nebula is an excellent template for low-UV illuminated PDRs \citep{Abergel2003}. The Horsehead is located at the western side of the Orion B molecular cloud and is viewed nearly edge-on with respect to the illumination source (see~Fig.~\ref{fig:hh-multiphase}), the massive O9.6V star $\sigma$ Ori, which is located $3.5$~pc away from the cloud \citep{Abergel2003,Schirmer2020}. The ionizing (EUV) photons arising from the star gave birth to the IC~434 \HII{} region, and the FUV photons have shaped the edge of the Orion B molecular clouds giving rise to multiple PDRs, most notably the top of the Horsehead. The Horsehead is illuminated by a moderate radiation field ($G_0 \sim 100$) and it has been studied extensively at different wavelengths. In particular, infrared H$_2$ observations together with low$-J$ CO isotopologues observed at millimeter wavelengths suggest that a steep density gradient must be present at the cloud edge \citep{Habart2005}. Furthermore, single-dish and interferometric observations have thoroughly characterized its molecular content, and have constrained the physical structure of the nebula at $0.01-0.02$~pc spatial scales \citep{Goicoechea2006,Pety2012,Gratier2013,Guzman2013,Guzman2014}. These studies, however, provide a low angular resolution ($5-10^{\prime\prime}$) view of the nebula.

Here we present ALMA observations at 0\farcs5 angular resolution (corresponding to 0.001~pc or 207~au at a distance of $\sim 400$~pc) of the CO and HCO$^+$ line emission at the surface layers of the iconic Horsehead nebula. We discuss their role as a tracers of the physical conditions of gas and provide constraints for parameters commonly used in PDR models. The observations and data reduction are described in Sect.~\ref{sec:Observations}, whereas the observational results are presented in Sect.~\ref{sec:ObsResults}. In Sect.~\ref{sec:PhysicalConditions} we describe the radiative transfer modeling of the molecular emission and derive constraints on the physical conditions of the gas. In Sect.~\ref{sec:PDRModels} we describe the PDR models that best reproduce the observations. We discuss the evidence of a steep density gradient and dynamical effects in the Horsehead in Sect.~\ref{sec:Discussion}, and we summarize the results and conclusions in Sect.~\ref{sec:Summary}.

\begin{figure*}[t!]
\centering
\includegraphics[width=0.9\linewidth]{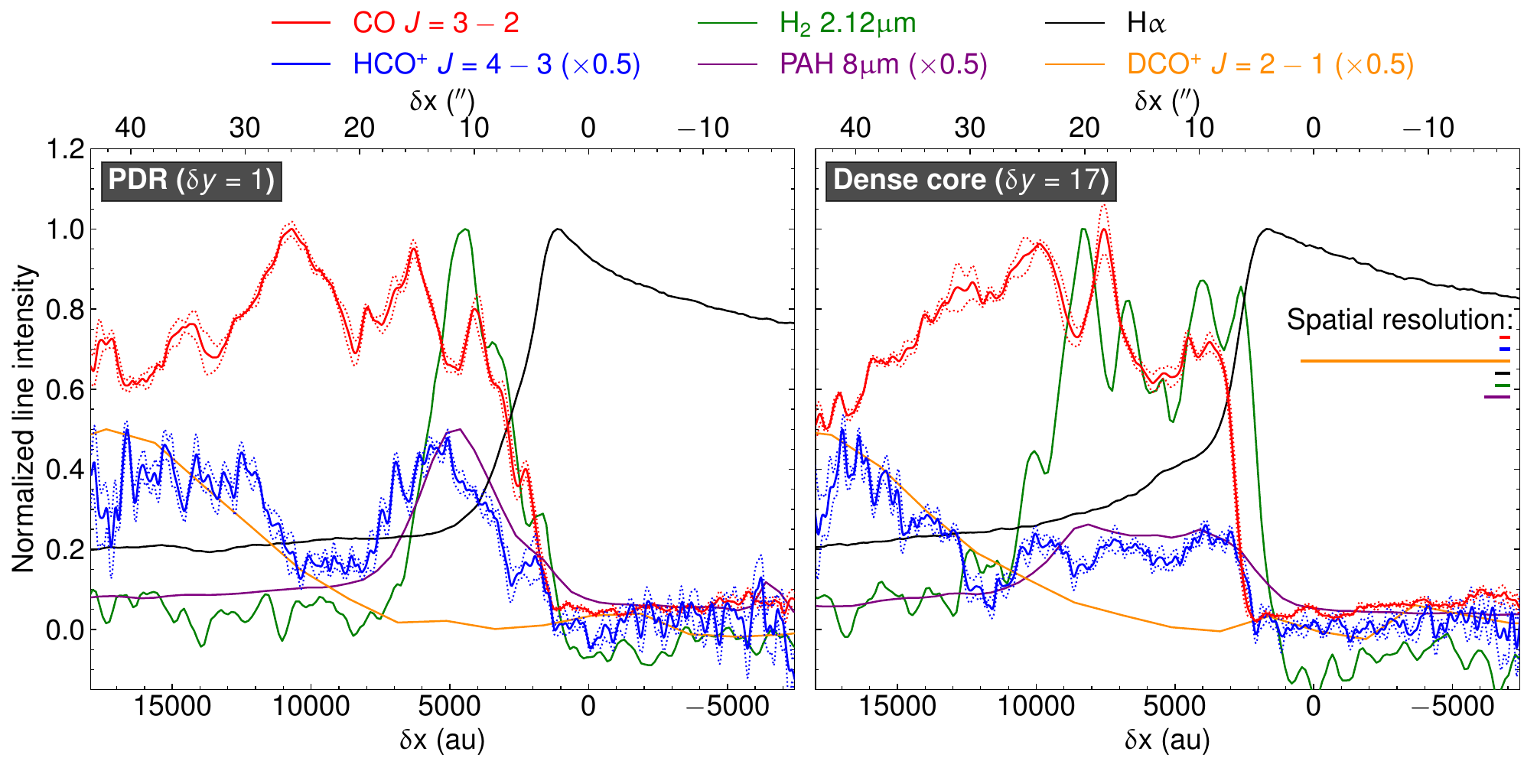}
\caption{Normalized integrated intensity profiles of the CO $J=3-2$ (solid red curve) and HCO$^+$ $J=4-3$ (solid blue curve) lines along the direction of the exciting star at the PDR (left) and dense core positions (right), averaged over $1^{\prime\prime}$ in the $\delta y$ direction (see right panel in Fig.~\ref{fig:hh-fronts}). The emission from other species such as DCO$^+$ $J=2-1$ \citep[orange curve,][]{Pety2007}, \Ha{} \citep[black curve,][]{Pound2003}, H$_2$ $2.12\mu$m \citep[green curve,][]{Habart2005}, and PAH $8\mu$m \citep[purple curve,][]{Schirmer2020} are also included. The spatial resolution of the observations is plotted in the top right corner. The dotted red and blue lines display the $\pm1\sigma$ significance levels for CO and HCO$^+$, respectively, corresponding to the standard deviation of the average in the $\delta y$ direction.}
\label{fig:hh-profiles}
\end{figure*}

\section{Observations}\label{sec:Observations}

The edge of the Horsehead nebula was observed with ALMA between October 7 2019 and December 12 2019, during Cycle 7 (2019.1.00558.S, PI: V.V. Guzmán). The observations were carried out in Band 7 using the 12~m array with baseline lengths between 15 and 500~m, the Atacama Compact Array (ACA) with baseline lengths between 8 and 44~m, and with the Total Power (single-dish). We observed a multiple-pointing mosaic with 39 and 14 fields, for the 12~m and ACA, respectively, both centered at $\alpha(2000) = 05^\mathrm{h}40^\mathrm{m}53^\mathrm{s}$; $\delta(2000) = -02^\circ27^\prime45^{\prime\prime}$, and covering a field-of-view (FoV) of approximately $50^{\prime\prime} \times 50^{\prime\prime}$. Two spectral windows, centered at $345.80$ and $356.73$~GHz, were included in order to observe the CO $J=3-2$ and HCO$^+$ $J=4-3$ lines, respectively. Other spectral windows were included in the setup targeting lines from reactive molecular ions (e.g., SH$^+$, SO$^+$, and HOC$^+$), but their analysis will be discussed in a future work.

The data reduction and imaging were performed using the Common Astronomy Software Application (CASA\footnote{https://casa.nrao.edu/}) version 5.6.1 \citep{2007ASPC..376..127M}. The different execution blocks were calibrated following the standard pipeline of The Joint ALMA Observatory (JAO). The calibrated visibilities of the 12~m array and ACA were merged into a single Measurement Set with the \code{concat} task, and then the data was imaged with the \code{tclean} function. The spectral cubes were cleaned using a Briggs weighting with a robust parameter of $0.5$. After trying different cleaning masks, we cleaned the images without any mask in order to include all the line emission from the molecular cloud and the \HII{} region, avoiding any possible biases\footnote{To help the cleaning process, we initially included a cleaning mask defined in two different ways. First, a mask was created manually by selecting regions with line emission in each channel. Then, we tried the auto-masking procedure implemented in CASA. However, in both cases the weak line emission present in the \HII{} region was filtered out.}. In addition, the multiscale CLEAN algorithm was used to deconvolve the dirty images in order to recover the different spatial scales of the emission. Finally, to recover the most extended emission, the Total Power observations were merged with the interferometric ones (corrected by the primary beam attenuation) using the \code{feather} function in CASA. After including the Total Power observations, the flux density measured in the molecular cloud increased by a factor of $\sim6$. This emphasizes the importance of using single-dish observations as "zero-" and "short-spacing" visibilities when imaging the extended emission of molecular clouds.

The final cubes have $\sim0.22$~km~s$^{-1}$ channel spacing and an average beam size of $0\farcs67 \times 0\farcs52$ (PA = $-75^\circ$). The achieved rms noise measured in line-free channels is $\sim5$~mJy~beam$^{-1}$, with a flux accuracy of approximately $10\%$. The integrated line intensity maps (see Fig.~\ref{fig:hh-grid}), as well as the peak brightness temperature map of the CO $J=3-2$ line (see Fig.~\ref{fig:hh-Tpeak}), were made with the \code{immoments} task in CASA. All the final products were rotated $14^{\circ}$ counterclockwise to bring the exciting star direction in the horizontal $\delta x$ direction. The observation parameters for each transition are detailed in Table~\ref{table:1}.

\section{Observational results}\label{sec:ObsResults}

\begin{figure*}[ht]
\centering
\includegraphics[width=0.9\linewidth]{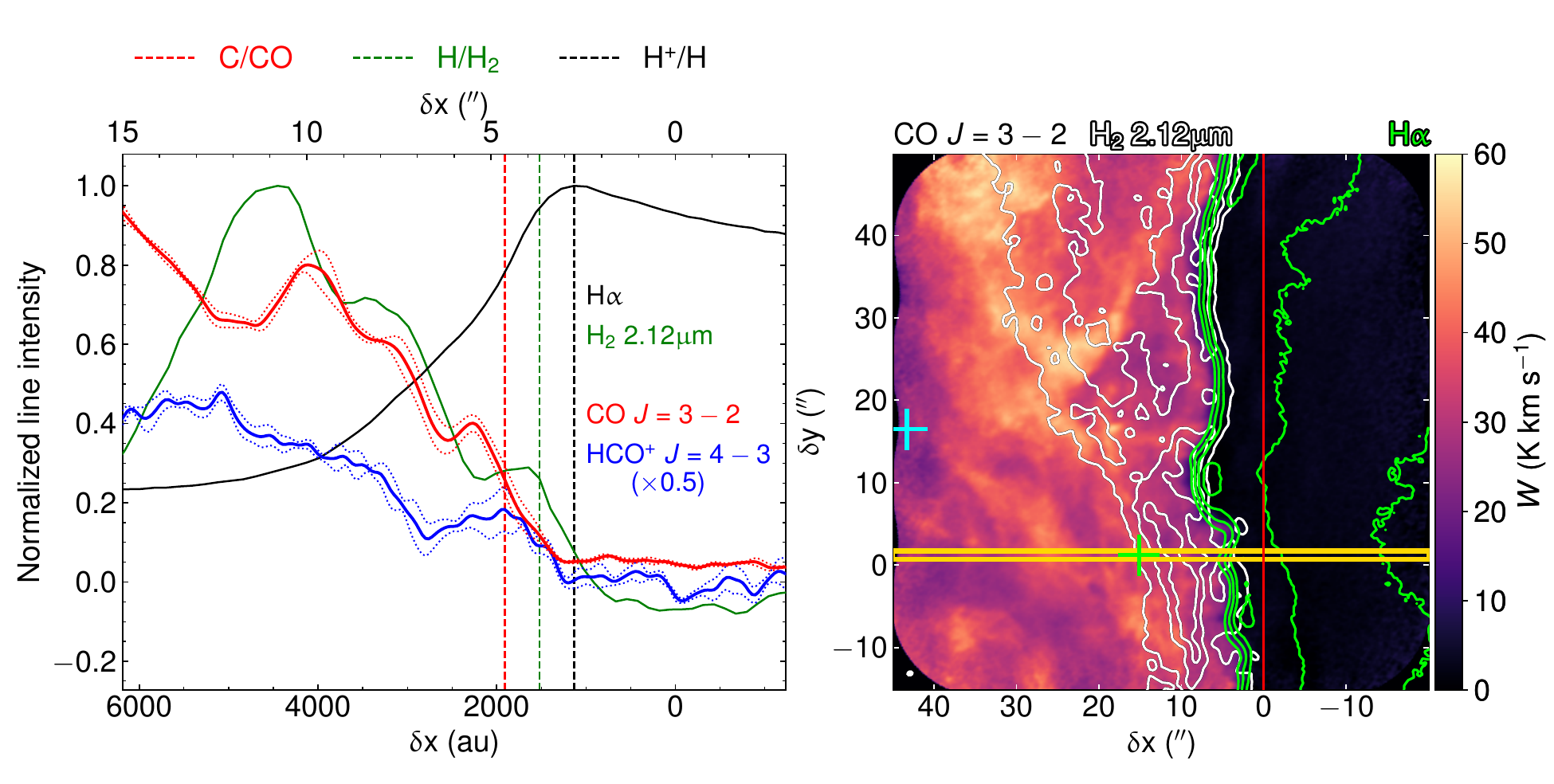}
\caption{Definition of the different transition zones. Left: Zoom in version of the normalized integrated intensity profiles shown in the left panel of Fig.~\ref{fig:hh-profiles} at the vertical position of the PDR, but only including the HCO$^+$ $J=4-3$ (solid blue curve), the CO $J=3-2$ (solid red curve), the \Ha{} (solid black curve), and the H$_2$ lines (solid green curve). The location of the ionization front, the dissociation front, and the C$^+$/C/CO transition zone are shown by the black, green, and red vertical dashed lines, respectively. The dotted red and blue lines display the $\pm1\sigma$ significance levels for CO and HCO$^+$, respectively, corresponding to the standard deviation of the average in the $\delta y$ direction. Right: CO $J=3-2$ integrated intensity map, showing the horizontal cut used to extract the intensity profiles at the vertical position of the PDR. The yellow rectangle shows the area where the emission of the different tracers was averaged, but in the zoom in version shown in the left panel this area was restricted between $\delta x = -3^{\prime\prime}$ to $15^{\prime\prime}$ for the purpose of defining the transition zones. The two crosses, the red vertical line, and the different contours are the same as in Fig.~\ref{fig:hh-grid}. The angular resolution is plotted in the bottom left corner.}
\label{fig:hh-fronts}
\end{figure*}

\begin{figure*}[ht]
\centering
\includegraphics[width=0.8\linewidth]{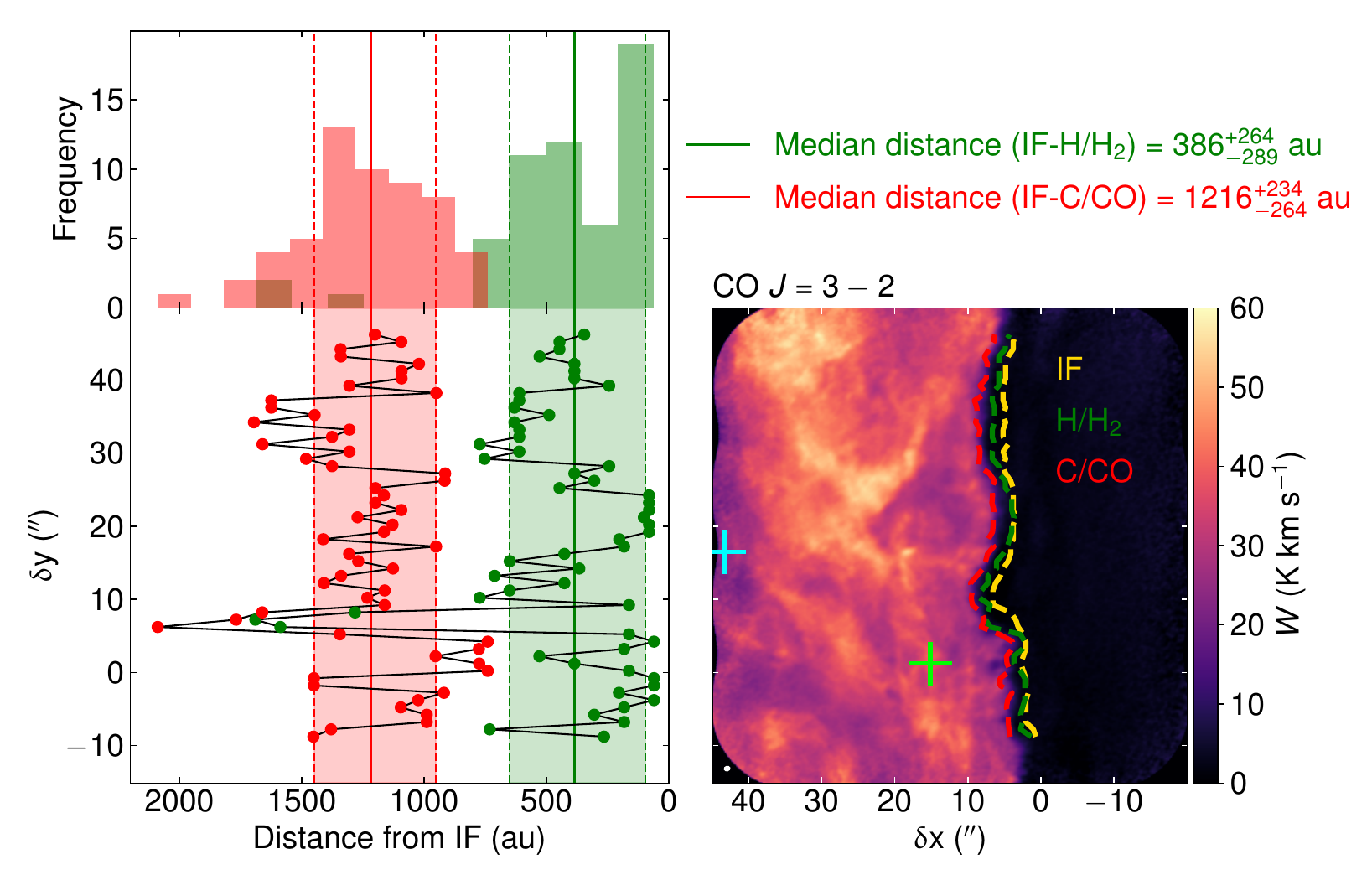}
\caption{Location of the ionization, dissociation, and C$^{+}$/C/CO transition zones. Left: Distances of the dissociation front (H/H$_2$, green dots) and C$^{+}$/C/CO transition zone (C/CO, red dots) with respect to the ionization front (IF) for different $\delta y$ vertical positions. The distributions of the distances are shown in the top panel, where the solid vertical lines represent the median values, and the dashed lines represent the uncertainties based on the 16th and 84th percentiles of each distribution. Right: IF (yellow), H/H$_2$ (green), and C/CO (red) transition zones overlaid on top of the CO $J=3-2$ integrated intensity map. The two crosses mark the PDR and dense core positions. The angular resolution is plotted in the bottom left corner.}
\label{fig:hh-distances}
\end{figure*}

\subsection{CO and HCO$^+$ ALMA maps}

We have mapped the emission from the CO $J=3-2$ and HCO$^+$ $J=4-3$ molecular lines at the edge of the Horsehead PDR at high angular resolution (0\farcs6). The integrated intensity maps are shown in Fig.~\ref{fig:hh-grid}, whereas the channel maps are presented in Appendix \ref{App:ChannelMaps}. The red vertical line represents the horizontal zero of the maps ($\delta x = 0$), which corresponds to the empirical PDR edge commonly defined from the sharp boundary traced by start of the H$_2$ emission \citep{Pety2005}. The two crosses represent the so-called PDR position ($\delta x \approx 15^{\prime\prime},\:\delta y \approx 1^{\prime\prime}$), commonly defined as the peak of the HCO line emission \citep{Gerin2009}, and the dense core position ($\delta x \approx 44^{\prime\prime},\:\delta y \approx 17^{\prime\prime}$), defined as the peak of DCO$^+$ $J=2-1$ line emission \citep{Pety2007}. With the new higher angular resolution ALMA observations we see that the location of the transition from atomic to molecular gas actually starts closer to the \HII{} regions than previously known. 

The velocity-integrated CO map reveals the structure of the molecular gas, which is characterized by the presence of bright structures embedded in an extended fainter emission pattern throughout the cloud (see top left panel in Fig.~\ref{fig:hh-grid}). The brightest structures form a web of connected filaments that are mostly parallel to the ionization front. The CO emission is present up to the edge of the cloud, and the right edge of the emission matches very well with the left side of the optical \Ha{} emission line, which arises from the ionized gas in the \HII{} region IC 434. The striking agreement between the edges of the CO and \Ha{} emission in the cloud indicates a very sharp transition from ionized to molecular gas, leaving a thin layer of neutral atomic gas in between. The thickness of this layer is estimated to be a few hundred au (see Fig.~\ref{fig:hh-multiphase} and Sect.~\ref{subsec:TransitionZones}). Likewise, the overlap between the CO and HCO$^+$ maps with the H$_2$ emission (white contours, Fig.~\ref{fig:hh-grid}) suggests that there is almost no CO-dark H$_2$ gas.

The HCO$^+$ line emission also shows some filamentary structure, but in contrast to CO, the emission is more concentrated in two vertical filaments (see bottom left panel in Fig. \ref{fig:hh-grid}). We find that HCO$^+$ traces two different environments: the cold dense gas in the FUV-shielded part of the cloud, and the warm gas in the FUV-exposed part of the cloud. To first order, the HCO$^+$ $J=4-3$ integrated line intensity ($W$ in K km s$^{-1}$) should scale with the gas density \citep{Goicoechea2016}, $n_{\mathrm{H}}$. Indeed, the peak of the HCO$^+$ line emission is close to the dense core position (cyan cross, Fig. \ref{fig:hh-grid}), which is characterized by low gas temperatures and high gas densities \citep[$T\approx20$~K and $n_{\mathrm{H}}\gtrsim 10^5$ cm$^{-3}$,][]{Pety2007}. However, HCO$^+$ is also bright at the edge of the cloud where the PDR is located (green cross, Fig. \ref{fig:hh-grid}). In contrast to the dense core, the PDR is characterized by the presence of warm UV-illuminated gas with relatively high densities \citep[$T\approx 60$ K and $n_{\mathrm{H}} \approx 6\times10^4$ cm$^{-3}$,][]{Gerin2009}. Therefore, the HCO$^+$ line emission not only traces cold dense gas, but also regions exposed to the FUV radiation field. Moreover, the HCO$^+$ filament at the PDR resembles the emission from the H$_2$ rovibrational line and the PAH emission at $8$ $\mu$m resolved by \textit{Spitzer} \citep{Schirmer2020} (see bottom panels in Fig.~\ref{fig:hh-grid}), which are sensitive to the UV radiation field and the gas density \citep{Habart2005}. Indeed, the PAH emission spatially coincides with the emission of H$_2$, which reinforces the previous expectations based on ISOCAM observations \citep{Abergel2003}, that is the mid-IR filament-shaped emission at the edge of the Horsehead nebula is mainly due to dense material illuminated by UV radiation seen nearly edge-on \citep{Habart2005}.

\subsection{Integrated intensity profiles}

To quantify the spatial distribution of the CO $J=3-2$ and HCO$^+$ $J=4-3$ line emission, we extracted horizontal cuts of the emission along the Horsehead PDR. For this, we averaged and normalized the integrated line intensity of the maps over $1^{\prime\prime}$ in the $\delta y$ direction at different vertical positions to resolve the variation of the emission across $\delta y$. To illustrate this, the profiles of two relevant vertical positions are shown in Fig.~\ref{fig:hh-profiles}: the dense core and the PDR\footnote{The PDR is located at $\alpha(2000) = 05^\mathrm{h}40^\mathrm{m}53.936^\mathrm{s}$; $\delta(2000) = -02^\circ28^\prime00^{\prime\prime}$ and the dense core is located at $\alpha(2000) = 05^\mathrm{h}40^\mathrm{m}55.61^\mathrm{s}$; $\delta(2000) = -02^\circ27^\prime38^{\prime\prime}$.}. 

We also extracted horizontal cuts of other tracers to determine the different transition zones and the physical conditions of the emitting gas as a function of the vertical position $\delta y$. In particular, we included observations from the \Ha{} line emission taken with the $0.9$m Kitt Peak National Observatory (KPNO) telescope \citep{Pound2003}, which trace the hot ionized gas in the \HII{} region adjacent to the Horsehead nebula with an approximate angular resolution of 1$^{\prime\prime}$. We also included observations from the H$_2$ $2.12$ $\mu$m rovibrational line, taken with the Son Of Isaac (SOFI) infrared imager-spectrometer at the New Technology Telescope (NTT) \citep{Habart2005}, which trace the molecular hydrogen excited by interactions with the UV radiation field at the edge of the nebula resolving also up to scales of $\sim$1$^{\prime\prime}$. The emission of the DCO$^+$ $J=2-1$ line observed by the IRAM-30m telescope \citep{Pety2007} was also included in order to trace the dense gas that is shielded from FUV photons. Finally, we included the emission from polycyclic aromatic hydrocarbons (PAHs) observed with the Infrared Array Camera (IRAC) from \textit{Spitzer} at $8\:\mu$m \citep{Schirmer2020}. PAHs have been found to be bright at the surface layers of the Horsehead, similar to the H$_2$ line \citep{Abergel2003,Habart2005}, and trace the atomic layer in other PDRs. The normalized profiles extracted from the other tracers at the vertical position of the PDR and dense core are also shown in Fig.~\ref{fig:hh-profiles}. An example of the region covered by a single horizontal cut is shown in the right panel of Fig.~\ref{fig:hh-fronts}, for the particular case of the PDR position.

The integrated intensity of both CO and HCO$^+$ lines increases rapidly from the edge of the nebula inward, similar to the H$_2$ line intensity (see Fig.~\ref{fig:hh-profiles}). Interestingly, in both cuts (PDR and dense core vertical positions), there is a small but visible shift between the rise of H$_2$ emission and the rise of CO emission, which indicates that we are effectively resolving the separation between the H/H$_2$ and C$^+$/C/CO transitions. The \Ha{} emission, however, is rather flat inside the \HII{} region and decreases inwards the molecular cloud but with a smaller slope. It is important to notice that the Horsehead edge is not viewed completely edge-on, but with a small inclination estimated to be $\sim6^{\circ}$ \citep{Habart2005}. We are therefore seeing a fraction of the surface of the cloud in these observations. 

\begin{figure*}[ht]
\centering
\includegraphics[width=0.9\linewidth]{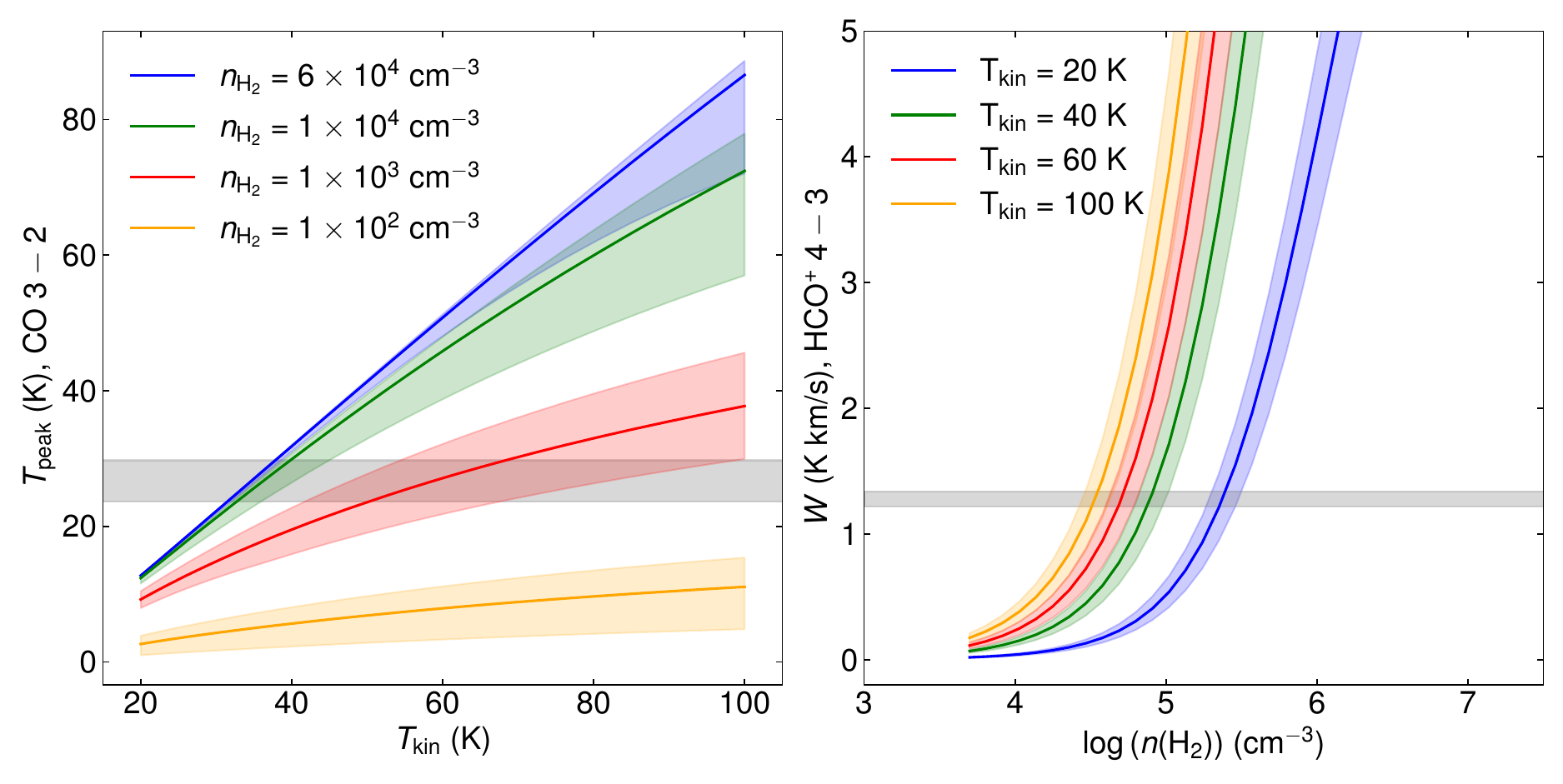}
\caption{Non-LTE radiative transfer models for different gas physical conditions at the PDR. Left: Expected CO $J=3-2$ peak brightness temperature as a function of the gas temperature for $N(\mathrm{CO})=(1.0-5.0) \times 10^{17}$ cm$^{-2}$ and different gas densities (colored lines). Right: Expected HCO$^+$ $J=4-3$ integrated line intensity as a function of the gas density for $N(\mathrm{HCO}^+) = (0.8-1.2) \times 10^{13}$ cm$^{-2}$ and different gas temperatures (colored lines). The solid lines display the model predictions for $N(\mathrm{CO})= 3 \times 10^{17}$ cm$^{-2}$ and $N(\mathrm{HCO}^+) = 1 \times 10^{13}$ cm$^{-2}$, and the colored areas represent the possible range of values for the assumed range of column densities. In both panels, the gray shaded areas represent the observations at the PDR position, with a $\pm 1\sigma$ uncertainty.}
\label{fig:hh-excitation-models}
\end{figure*}

\subsection{Transition zones}\label{subsec:TransitionZones}

The CO $J=3-2$ integrated line intensity can be used to locate the region where CO starts to form, and hence of the C$^+$/C/CO transition zone (or C/CO, in short), which can be compared with the position of the ionization front (H$^+$/H transition traced by \Ha{}) and the dissociation front (H/H$_2$ transition traced by H$_2$). The spatial coincidence between the line emission of CO and H$_2$ at the surface layers of the molecular cloud implies that the H/H$_2$ and C/CO transition zones are spatially very close, being apparently separated by no more than $\sim 1200$~au (approximately $3^{\prime\prime}$ in Fig.~\ref{fig:hh-grid}). A similar finding was previously observed in the case of the Orion Bar PDR \citep{Goicoechea2016}, which contradicts historical constant density PDR models that commonly predict the C/CO transition to be located significantly deeper inside the cloud compared to the H/H$_2$ transition \citep{Hollenbach1999}. 

Based on the \Ha{}, H$_2$, and CO emission profiles we quantified the location of the different transition zones, by assuming a fully edge-on geometry. The ionization front is defined as the location of the peak of the \Ha{} emission and the dissociation front is defined as the location where the derivative of the H$_2$ profile is maximum. Similarly, the C/CO transition was defined as the location where the slope of the CO profile reaches its maximum value. It is important to keep in mind that the above definitions are rough approximations of the different transition zones, especially for the ionization front, since the \Ha{} profile is much smoother than in the cases of H$_2$ and CO emission, which follow very similar sharp rises with a spatial shift that is easy to define. Therefore, the accuracy of the position of the ionization front is probably not as good as in the case of the H/H$_2$ and C/CO transition zones, but it is still a reasonable approximation.

Figure~\ref{fig:hh-fronts} shows the relevant profiles and the location of the different transition zones at the vertical position of the PDR. Additionally, to compare with predictions from PDR models, we estimated the distance projected in the plane of the sky of the dissociation front and C/CO transition zone, both with respect to the position of the ionization front. Replicating the same procedure but now for the profiles extracted at different vertical positions, we were able to derive the variation of the transition zones as a function of $\delta y$, as shown in Fig.~\ref{fig:hh-distances}.

We grouped the distance values into distributions (see top left panel, Fig.~\ref{fig:hh-distances}), and based on the 16$^{\mathrm{th}}$, 50$^{\mathrm{th}}$, and 84$^{\mathrm{th}}$ percentiles, we determined the median value and the uncertainties for the distance of the different transition zones, which are illustrated by the colored areas in the bottom left panel of Fig.~\ref{fig:hh-distances}. In the case of the dissociation front (green histogram, Fig.~\ref{fig:hh-distances}) there is a bimodal distribution, with a clear peak around $\sim100$~au and another less prominent one centered at $\sim500$~au. Taking the median of the distribution, we estimate that a representative value of the distance between the H$^+$/H and H/H$_2$ transition zones (defined as the atomic layer) is $386^{+264}_{-289}$~au, where the two modes of the distribution mentioned above are included in the uncertainties. Similarly, we estimate the separation between the H$^+$/H and C/CO transition zones to be $1216^{+234}_{-264}$~au (red histogram, Fig.~\ref{fig:hh-distances}). If we take into account the Horsehead PDR is not fully edge-on, then the real distances should be slightly smaller. Still, considering the small inclination angle, we do not expect these projected distances to change significantly. However, given the uncertainties in the estimates of the front positions, and taking into account that a significant fraction of the measured distances between the H$^+$/H and H/H$_2$ transitions is around $\sim100$~au, we consider that our measurement of the length of the atomic layer should be interpreted only as an upper limit, as with ALMA we can resolve physical scales up to $\sim200$~au.

Based on the derived distances, we can estimate the width of the CO-dark H$_2$ gas (DG) layer between the H/H$_2$ and C/CO transition zones. According to the median value of each distribution, the size of the DG layer is estimated to be $\ell_{DG} \sim 830$~au $=1.24\times 10^{16}$~cm. The above implies a gas column density of about $N_{\mathrm{H}} \sim (0.12-4.96)\times 10^{20}$~cm$^{-2}$, assuming a gas density between $n_{\mathrm{H}} = 10^{3}-4\times10^{4}$~cm$^{-3}$, where the minimum and maximum values correspond to estimates from our ALMA CO observations (see Sect.~\ref{subsec:DensGrad} for more details) and from previous C$^+$ observations \citep{Pabst2017}, respectively. Assuming the relation $N_{\mathrm{H}} = 1.9 \times 10^{21} A_{\mathrm{V}}$~cm$^{-2}$ for the local Galaxy, then the width of the DG layer in terms of visual extinction should be $A_{\mathrm{V}} = 0.006 - 0.26$~mag, which is lower than the $A_{\mathrm{V}} = 0.6 - 0.8$~mag value estimated by previous DG models, although at slightly lower $G_0$ values \citep{Wolfire2010}. Therefore, our results are consistent with a very thin atomic layer and demonstrate that there is a small amount of CO-dark H$_2$ gas between the molecular CO and the ionized gas in the Horsehead nebula.

\section{Physical conditions of the gas}\label{sec:PhysicalConditions}

The peak brightness temperature ($T_{\mathrm{peak}}$ in K) of a line can be approximated as $T_\mathrm{peak} = [J(T_{\mathrm{ex}}) - J(T_{\mathrm{bg}})](1-e^{-\tau_{\mathrm{line}}}) \approx J(T_{\mathrm{ex}})(1-e^{-\tau_{\mathrm{line}}}) = E_{\mathrm{up}}/k_{B} \times \left[\exp\left(E_{\mathrm{up}}/k_{B}T_{\mathrm{ex}}\right)-1\right]^{-1}(1-e^{-\tau_{\mathrm{line}}})$, where $T_{\mathrm{ex}}$ and $T_{\mathrm{bg}}$ are the excitation and background temperature, respectively ($T_{\mathrm{ex}} \gg T_{\mathrm{bg}}$ for CO in warm gas), $E_\mathrm{up}$ is the upper level energy for a given transition, and $\tau_{\mathrm{line}}$ is the optical depth of the line. However, if the line is optically thick, then we have $(1-e^{-\tau_{\mathrm{line}}}) \rightarrow 1$, and thus $T_{\mathrm{peak}}$ only depends on $T_{\mathrm{ex}}$. Therefore, the peak brightness temperature of optically thick lines ($\tau_{\mathrm{line}} \gg 1$) is a powerful tracer of the gas temperature, $T_{\mathrm{kin}}$, provided that the gas density is well above the critical density of the line in question ($n_{\mathrm{H}} \gg n_{\mathrm{cr}}$). In particular, CO transitions corresponding to low rotational levels are optically thick ($\tau_{\mathrm{CO}3-2} \gg 1$) and have relatively low critical densities, falling between $n_{\mathrm{cr}} = (0.22 - 3.73) \times 10^4$~cm$^{-3}$ for the first three radiative transitions. Consequently, for gas densities above $\sim 10^{4}$~cm$^{-3}$, the low-$J$ CO lines should be thermalized and \TpeakCO{} approximate the excitation temperature \citep{Goicoechea2016}.

On the other hand, the integrated intensity of the HCO$^+$ $J=4-3$ line (\WHCOp{}) can be used to trace the gas density if the line opacity is low, since it has a relatively high critical density ($n_{\mathrm{cr}} > 5 \times 10^{6}$ cm$^{-3}$) compared to typical gas density values found in the Horsehead. In particular, when there is subthermal excitation, \WHCOp{} is approximately linearly proportional to the column density, which can be written as a function of the gas density as $N(\mathrm{HCO}^+) = x(\mathrm{HCO}^+)n_{\mathrm{H}}\ell$, where $\ell$ is the cloud length along the line of sight. Therefore, for $n_{\mathrm{H}} < n_{\mathrm{cr}}/\tau_{\mathrm{HCO}^+4-3}$ the line is subthermally excited, and \WHCOp{} should be linearly proportional to $n_{\mathrm{H}}$ \citep{Goicoechea2016}.

\begin{figure*}[ht]
\centering
\includegraphics[width=1.0\linewidth]{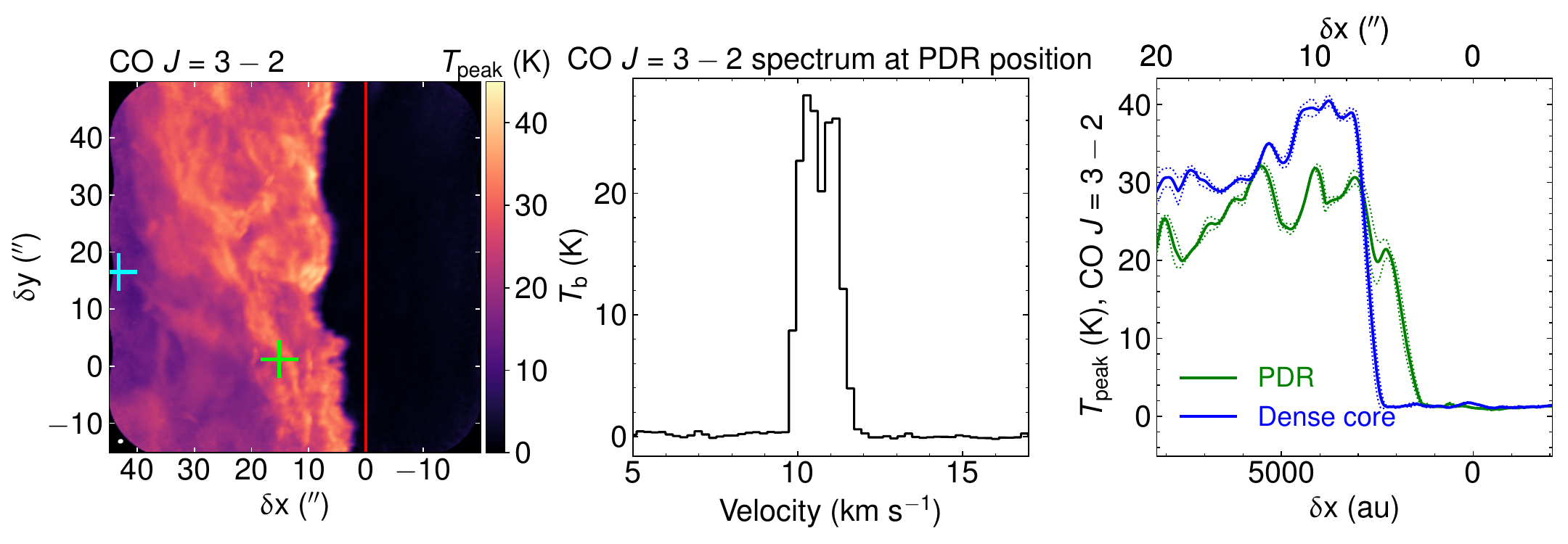}
\caption{CO $J=3-2$ peak brightness temperature. Left: \TpeakCO{} map rotated $14^\circ$ counterclockwise to bring the illuminating star direction in the horizontal direction. The two crosses and the red vertical line are the same as in Fig.~\ref{fig:hh-grid}, and the angular resolution is plotted in the bottom left corner. Middle: Spectrum of the CO $J=3-2$ line at the PDR position. Right: \TpeakCO{} profiles at the vertical position of the PDR (solid green curve) and dense core (solid blue curve), averaged over $1^{\prime\prime}$ in the $\delta y$ direction. The dotted green and blue lines display the $\pm1\sigma$ significance levels, corresponding to the standard deviation of the average in the $\delta y$ direction.}
\label{fig:hh-Tpeak}
\end{figure*}

\subsection{Radiative transfer models}

To assess the physical conditions (temperature and density) of the emitting gas at the edge of the Horsehead nebula, we ran a grid of models using the nonlocal thermodynamic equilibrium (non-LTE) radiative transfer code RADEX\footnote{https://home.strw.leidenuniv.nl/moldata/radex.html} \citep{vanderTak2007}. We explored a range of gas temperatures and densities, with fixed CO and HCO$^+$ column densities, and compare the observed \TpeakCO{} and \WHCOp{} values with those predicted by RADEX. We included the inelastic collisional rates of CO with both H \citep{Walker2015} and H$_2$ \citep[with ortho- and para-H$_2$ distinction,][]{Yang2010}, and of HCO$^+$ with H$_2$ \citep{Flower1999}. In the case of HCO$^+$, collisional rates are also available for electrons \citep{Faure2001} but they do not have a strong impact on the results even though the ionization fraction is high in the Horsehead PDR \citep{Goicoechea2009} and, therefore, they were not included in the models. Alternatively, collisions with atomic H can have a meaningful contribution in PDRs since the gas is not fully molecular ($f<1$, where $f=2n(\mathrm{H}_2)/[n(\mathrm{H})+2n(\mathrm{H}_2)]$ is the molecular gas fraction). We tried a range of values for the molecular fraction ($f=0.5-1$), and found that the results did not change significantly. We therefore adopted a value of $f=0.8$, similar to other PDRs \citep{Goicoechea2016}. The temperature of the cosmic microwave background was fixed at $T_{\mathrm{CMB}} = 2.73$~K and we assumed a line width of $\mathrm{FWHM} = 1$~km~s$^{-1}$ based on the observed line profiles. We ran our grid of models varying the H$_2$ density from $5\times10^3$ to $1 \times 10^{7}$~cm$^{-3}$, with a fixed gas temperature when modeling the HCO$^+$ line, and varying the temperature from $20$ to $100$ K for fixed values of $n_{\mathrm{H}_2}$ when modeling the CO line. 

Regarding the column densities, they were fixed based on previous estimates. In the case of HCO$^{+}$, a column density of about $N(\mathrm{HCO}^{+}) \sim 8 \times 10^{12}$~cm$^{-2}$ was determined from multi$-J$ observations with IRAM-30m telescope in the Horsehead (M.G. Santa-Maria, priv. comm.). 
To take into account possible beam dilution effects, since our observations have higher angular resolution, we ran our models with a range of column densities that include larger values of $N(\mathrm{HCO}^+) = (0.8-1.2) \times 10^{13}$~cm$^{-2}$. In the case of CO, a column density of about $N(\mathrm{CO}) \sim 10^{18}$~cm$^{-2}$ was found at the PDR position based on C$^{18}$O observations with the IRAM Plateau de Bure Interferometer (PdBI) \citep{Pety2005} and the standard isotopic ratio of $^{16}\mathrm{O}/^{18}\mathrm{O}=560$ \citep{Wilson1994}. We used a slightly lower range of column densities of $N(\mathrm{CO})=(1-5) \times 10^{17}$~cm$^{-2}$. Indeed, previous C$^+$ observations show an extended emission from the neutral atomic layer \citep{Pabst2017,Bally2018} that coincides with part of the CO emission. Thus, part of the carbon inventory in the molecular cloud may be in its atomic form (either in C or C$^+$), and therefore the column density of CO could be lower than that estimated from C$^{18}$O, in particular at the cloud edge. The model predictions for \TpeakCO{} and \WHCOp{} are shown in Fig.~\ref{fig:hh-excitation-models}.

\subsection{Gas temperature}\label{subsec:Temperature}

We find that \TpeakCO{} is very sensitive to changes in $T_{\mathrm{kin}}$ as was expected (see left panel in Fig.~\ref{fig:hh-excitation-models}) and is not very sensitive to the density when the gas density is high ($\gtrsim 4\times10^4$~cm$^{-3}$). However, as soon as the gas density drops below $10^4$~cm$^{-3}$, \TpeakCO{} is no longer a good tracer of the gas temperature since the line is no longer thermalized. Our ALMA observations show that \TpeakCO{} $\lesssim 45$ K at the Horsehead edge (see left panel, Fig.~\ref{fig:hh-Tpeak}), however, the observed value at the PDR position is only \TpeakCO{} $\approx 30$ K ($\delta x \approx 15^{\prime\prime}$ in right panel, Fig.~\ref{fig:hh-Tpeak}). To reproduce the observed \TpeakCO{}, the gas temperature would have to be $\lesssim40$~K if we consider the gas density of $6\times10^4$~cm$^{-3}$, which has been previously inferred from lower angular resolution observations (blue line, left panel in Fig.~\ref{fig:hh-excitation-models}). This is lower than what has been previously estimated, also from low angular resolution observations, which suggests temperatures of about 60~K at the PDR position \citep{Pety2005}. We note that the ALMA single-dish observations presented here of the CO $J=3-2$ line are consistent with previous observations from Caltech Submillimeter Observatory (CSO) and the Atacama Pathfinder Experiment (APEX) telescope \citep{Philipp2006,Bally2018}. 

To understand the lower than expected gas temperature, we inspected the line profiles at different positions in the observed field of view. We found that near the PDR position the line profiles are not Gaussian, show asymmetries, and present evidence of self-absorption at the source velocity ($\sim10.5$~km~s$^{-1}$) that can be associated with CO-emitting layers with lower densities, suggesting the presence of a density gradient in the line of sight. The middle panel in Fig.~\ref{fig:hh-Tpeak} shows the extracted spectra at the PDR position, where a dip in emission is clearly seen at the line center. The same behaviour has been previously reported for the CO $J=1-0$ and $J=2-1$ lines with both single-dish \citep{Abergel2003} and interferometric \citep{Pety2005} observations. A similar effect has been found in the anomalous HCN~$J=1-0$ hyperfine line ratios toward the Horsehead PDR, which has been attributed to self-absorption produced by resonant scattering through a diffuse cloud envelope \citep{Goicoechea2022a}. Indeed, considering that the Horsehead is not fully edge-on, the presence of density gradients along the line of sight cannot be discarded. Therefore, we conclude that the gas temperatures inferred from the observed \TpeakCO{} values should be interpreted only as lower limits, since the CO emission could be affected by low density gas in the line of sight.

\subsection{Gas density}\label{subsec:Density}

We find that \WHCOp{} is indeed proportional to $n_{\mathrm{H}}$ with a characteristic behavior for a linear dependence on a linear-log plot (see right panel in Fig.~\ref{fig:hh-excitation-models}, where the $x$-axis is on logarithmic scale). The observed HCO$^+$ $J=4-3$ integrated line intensity is about $W_{4-3}^{\mathrm{HCO}^+} \approx 1.28 \pm 0.06$ K km s$^{-1}$ at the PDR position (located at $\delta x \sim 15^{\prime\prime}$, see Fig.~\ref{fig:hh-profiles}). In order to reproduce the observed \WHCOp{}, our models predict a gas density of $n_{\mathrm{H}} = (3.9-6.6) \times 10^4$~cm$^{-3}$ for a gas temperature of $T_{\mathrm{kin}} = 60$~K (red line, right panel in Fig.~\ref{fig:hh-excitation-models}), whereas for the lower limit that we obtained from \TpeakCO{} in the PDR ($T_{\mathrm{kin}} \gtrsim 40$~K), we estimate an upper limit for the gas density of about $n_{\mathrm{H}} \lesssim 10^5$~cm$^{-3}$ (green line, right panel in Fig.~\ref{fig:hh-excitation-models}). This is in good agreement with the estimated value from lower angular resolution observations \citep{Gerin2009} of $n_{\mathrm{H}} = 6 \times 10^4$~cm$^{-3}$, suggesting the PDR would not be strongly compressed compared to the ambient cloud. Interestingly, the HCO$^+$ integrated intensity is particularly bright in the surface layers of the Horsehead PDR compared to its surroundings (see Fig.~\ref{fig:hh-grid} and profiles in Fig.~\ref{fig:hh-profiles}). More precisely, there is a "gap" in the emission between the PDR and dense core positions ($\delta x \sim 25-30^{\prime\prime}$ in our figures), which could be a direct sign of compression at the edge of the Horsehead nebula, if we assume that \WHCOp{} is proportional to $n_{\mathrm{H}}$. 

To quantify the compression factor, we estimated the gas density in the gap. As one moves deeper into the molecular cloud toward the dense core position, $T_{\mathrm{kin}}$ is expected to decrease \citep{Pety2007,Gratier2013} while $N(\mathrm{HCO}^+)$ should increase \citep{Goicoechea2009}. Thus, by running RADEX with conservative physical conditions expected for the gas between the PDR and the dense core (namely, for $\delta x \sim 25-30^{\prime\prime}$ we assumed intermediate values of $T_{\mathrm{kin}} = 20 - 40$~K and $N(\mathrm{HCO}^+) = 5\times10^{13}$~cm$^{-2}$), and assuming a FWHM = $0.7$~km~s$^{-1}$ based on the observations, then the gas density that best reproduces the observed $W_{4-3}^{\mathrm{HCO}^+} = 0.55 \pm 0.15$~K~km~s$^{-1}$ integrated intensity in the gap is $n_{\mathrm{H}} = \left(0.5 - 2.4\right)\times 10^4$~cm$^{-3}$. By comparing the densities derived from \WHCOp{} for the PDR and the gap, we estimate a compression factor between $2-13$, which is in the lower range compared to other strongly illuminated PDRs, like the Orion Bar \citep[compression factor of about $5-30$,][]{Goicoechea2016}. The difference could be related to the order of magnitude difference in their respective UV radiation fields. Therefore, the gas at the Horsehead PDR is apparently not as highly compressed, which is consistent with the rather smooth HCO$^+$ line emission observed near the PDR position (see the bottom left panel in Fig.~\ref{fig:hh-grid}).

\begin{figure*}[ht]
\centering
\includegraphics[width=0.8\linewidth]{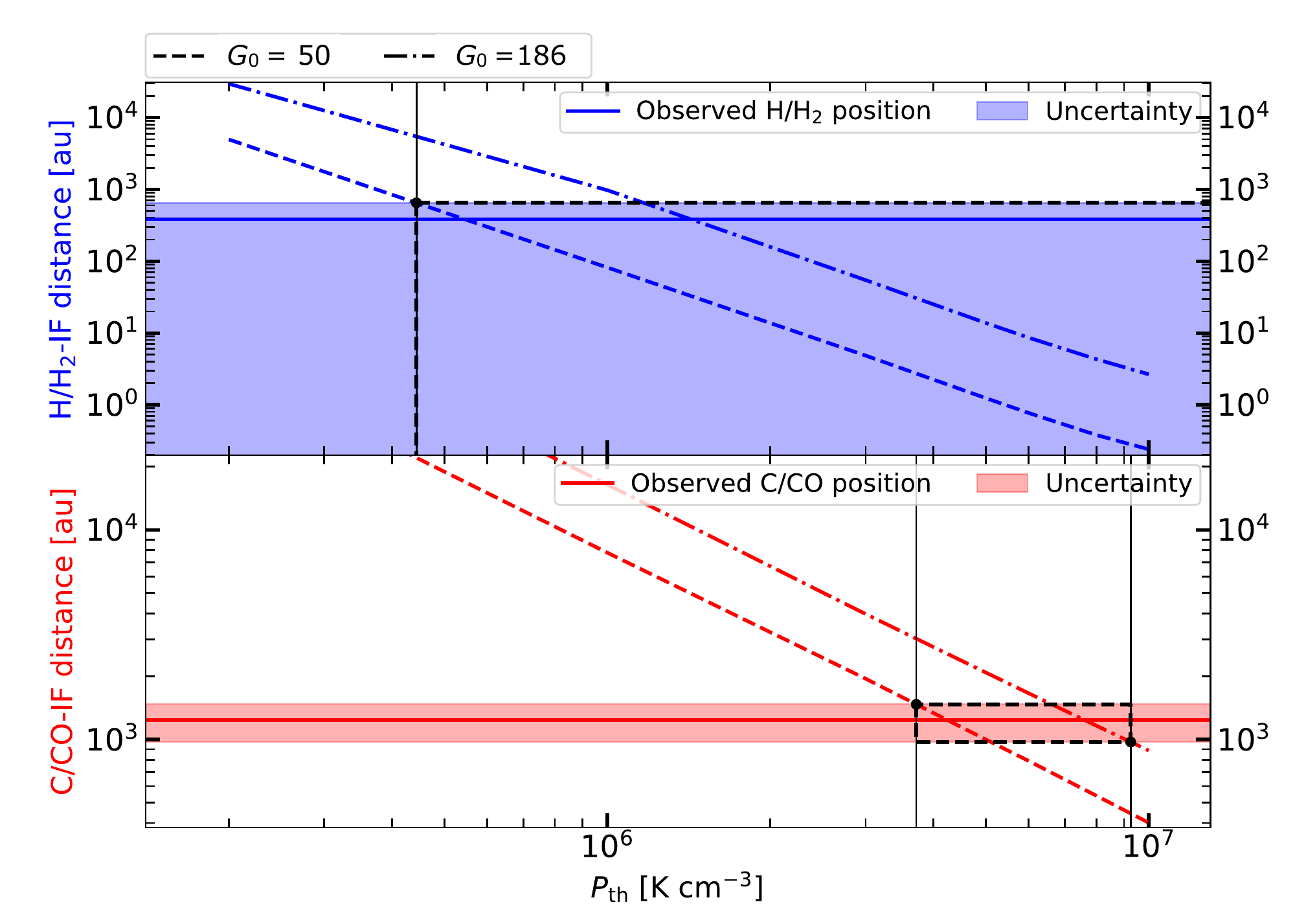}
\caption{Range of parameters compatible with the distances measured in the observations. Top: Distance between the ionization front (IF) and the H/H$_2$ transition (dissociation front). The blue zone represents the upper limit that has been derived for the distance between the IF and the dissociation front. The blue dashed lines indicates the distance found by the Meudon PDR code for two radiation fields ($G_0 = 50$ and $186$) and different thermal pressures ($P_{\mathrm{th}}$). The models compatible with the observations are those in the blue zone. Bottom: same but for the distance between the IF and C/CO transition. The domain of parameters that reproduce the distances is represented by the black dashed squares.}
\label{fig:hh-CompatibleModels}
\end{figure*}

\subsection{Gas pressure}\label{subsec:Pressure}

To further investigate the gas compression at the Horsehead PDR, we compared the thermal and nonthermal components of the gas pressure at different locations. In the case of the PDR position ($\delta x \sim 15^{\prime\prime}$), our estimates for the gas density, together with previous temperature measurements, would imply a thermal pressure of $P_{\mathrm{th}} = n_{\mathrm{H}} T_{\mathrm{kin}} \approx \left(2.3 - 4.0 \right) \times 10^6$~K~cm$^{-3}$ that is consistent with previous estimates \citep{Habart2005}. On the other hand, the assumed temperatures and the estimated range of densities for the gap imply thermal pressures of about $P_{\mathrm{th}} \approx \left(2.0-4.8\right)\times 10^{5}$~K~cm$^{-3}$, which are lower than the values at the PDR position. Instead, the above values agree with the thermal pressures predicted at the adjacent \HII{} region IC 434 \citep{Bally2018}, where $P_{\mathrm{th,\mathrm{\HII{}}}} \approx n_{e}T_{e} =\left(2.4-8.0\right)\times 10^{5}$~K~cm$^{-3}$. This difference in thermal pressures between the PDR position and its surroundings is consistent with a compression scenario.

Furthermore, based on the gas density values derived from HCO$^+$, along with the observed line profiles, we also estimated the turbulent ram pressure at different positions. This nonthermal component of the pressure could be defined as $P_{\mathrm{nth}} = \rho \sigma_{\mathrm{nth}}^2$, where the turbulent (nonthermal) velocity dispersion ($\sigma_{\mathrm{nth}}$) can be measured from the observed HCO$^+$ line profiles as $\sigma_{\mathrm{nth}}^2 = \sigma_{\mathrm{obs}}^2 - \sigma_{\mathrm{th}}^2$, where $\sigma_{\mathrm{th}}^2 = k T_{\mathrm{kin}}/\mu m_{\mathrm{H}}$. At the PDR position, we measured $\Delta v_{\mathrm{FWHM}} = 2\sqrt{2\ln2} \times \sigma_{\mathrm{obs}}\approx 0.95$~km~s$^{-1}$, which for $T_{\mathrm{kin}} = 60$~K implies a nonthermal pressure of approximately $P_{\mathrm{nth}} \approx 10^6$~K~cm$^{-3}$. Likewise, close to the gap of HCO$^+$ we obtained $\Delta v_{\mathrm{FWHM}} \approx 0.71$~km~s$^{-1}$, which for $T_{\mathrm{kin}} = 30$~K leads to $P_{\mathrm{nth}} \approx 10^5$~K~cm$^{-3}$. In both positions, we find that $P_{\mathrm{th}} \sim P_{\mathrm{nth}}$, which implies that the different pressure components are in equipartition.

\subsection{Estimated magnetic field strength}

If we extend the equipartition between thermal and turbulent (nonthermal) components of the pressure to the magnetic field component ($P_{B} = B^2/8\pi k$), then we can estimate the magnetic field strength in the Horsehead. By assuming that the magnetic pressure at the PDR position is on the order of $P_{B}^{\mathrm{PDR}} \sim 10^6$~K~cm$^{-3}$, then we infer a magnetic field strength of about $B \sim 60$~$\mu$G in that part of the molecular cloud. Our estimation is in good agreement with previous measurements in other sources of the Orion B molecular cloud, such as NGC 2024 \citep{Crutcher1999,Roshi2014} and NGC 2071 \citep{Matthews2002}, and with observations from polarized dust taken with the James Clerk Maxwell Telescope (JCMT) that recently deduced a magnetic field strength of $B=56 \pm 9$~$\mu$G in the Horsehead PDR \citep{Jihye2023}, indicating that the equipartition assumption is reasonable for the Horsehead conditions.

\section{PDR modeling}\label{sec:PDRModels}

To assess whether the extremely small separations that we observe between the ionization front (IF), the dissociation front (H/H$_2$), and the C$^+$/C/CO transition (C/CO) are consistent with the traditional stationary view of PDRs or are actually evidence of important dynamical effects, like photoevaporation as predicted by models \citep{Maillard2021}, we compare the measured distances between the fronts to the predictions of the Meudon PDR code\footnote{https://pdr.obspm.fr/} \citep{LePetit2006}. The Meudon PDR code computes the radiative transfer in UV lines and thus the photo-destruction rates of H$_2$ and CO, as well as their formation rates on grains and in the gas \citep{Goicoechea2007, LeBourlot2012}. The chemical abundances of these species in the cloud can then be determined as well as the locations of the H/H$_2$ and C/CO transitions. We use as observational constraints a distance of $1216^{+234}_{-264}$ au between the IF and C/CO transition, and an upper limit of $<650$ au for the distance between the IF and H/H$_2$ dissociation front (see Fig.~\ref{fig:hh-distances}). We use an upper limit for the latter distance since after considering the uncertainties the distance is comparable to the beam size of the CO observations.

We explore a range of radiation field intensities $G_0 = 50-186$ in Habing units (or, equivalently, $\chi = 30-110$ in Draine units), in agreement with previous studies of the Horsehead \citep{Habart2005}. As a first approximation for dust extinction, we use the standard Galactic extinction curve \citep{Fitzpatrick1986}. Particularly, for the Horsehead conditions, the absorption of dissociating photons is dominated by H$_2$ electronic absorption lines and not dust extinction \citep{Sternberg2014}, so that the choice of dust extinction properties is not expected to affect the position of the dissociation front significantly. For each produced PDR model, we determine the locations of the H/H$_2$ and C/CO transitions and compare their distances from the IF to the ones deduced from the observations. With the Meudon PDR code, we can simulate a medium with constant density or with constant thermal pressure (isobaric case). Alternatively, it can also read user defined density profiles. In the Horsehead PDR case, the presence of a strong density gradient had already been suggested by lower angular resolution observations \citep{Habart2005,Guzman2011}, where different prescriptions for a density profile were proposed. For instance, using H$_2$ and low$-J$ CO observations, a steep density gradient was found at the edge of the Horsehead and suggested a density profile represented as a power law followed by a constant density medium at $n_\mathrm{H} = 2\times10^5$ cm$^{-3}$ \citep{Habart2005}. A similar density profile was then used to reproduce observations of H$_2$CO \citep{Guzman2011}. Indeed, we verified that the PDR models with constant density cannot reproduce the observations. However, the comparison to the observed distances also shows that PDR models using previous density profile prescriptions are incompatible with the observations. The H/H$_2$ transition is too deep in the PDR by at least a factor 2 compared with the upper limit found with the observations, and the C/CO transition is too far by at least a factor 3.

These profiles were initially proposed to emulate the density gradient that counterbalances the decreasing temperature gradient in an isobaric PDR. Indeed, recent ALMA and \textit{Herschel} observations of molecular tracers at the edge of PDRs strongly suggest that the warm layer of PDRs is isobaric with relatively large thermal pressures controlled by the intensity of the incident dissociating radiation field \citep{Marconi1998, Goicoechea2016, Joblin2018, Wu2018, Bron2018, Maillard2021}. We thus also compared the measured front distances to isobaric PDR models. In this case, at each position the code solves the thermal balance and deduces the gas temperature based on the given thermal pressure and radiation field intensity. Since $P_{\mathrm{th}} = n_{\mathrm{H}}T_{\mathrm{kin}}$, then the density is derived at each position, creating a natural gradient of the density \citep{Joblin2018}. We produced a grid of PDR models with a range of pressures between $P_{\mathrm{th}}= 2 \times 10^5$ to $10^7$ K cm$^{-3}$, based on our estimations (see section~\ref{subsec:Pressure}). Figure~\ref{fig:hh-CompatibleModels} summarizes the range of parameters for which the models are found to be compatible with the observations presented in this paper. Models with $P_{\mathrm{th}}$ between $(3.7-9.2)\times10^{6}$ K cm$^{-3}$ can reproduce the observational constraints on the distances between the fronts, which is also consistent with our estimates for the thermal gas pressure. We also checked that these models are compatible with the intensity of the CO $J=3-2$ line. This demonstrates that it is possible to explain the short distances between the IF, the H/H$_2$, and C/CO transition zones in the frame of stationary models.

\section{Discussion}\label{sec:Discussion}

\subsection{Steep density gradient}\label{subsec:DensGrad}

The ALMA images imply a narrow neutral atomic layer at the surface of the Horsehead nebula ($\lesssim 600$~au). This intermediate zone between the \HII{} region and the molecular gas appears to be much thinner than in the case of the Orion Bar PDR, where the separation between the ionization and dissociation fronts is about $6000$~au \cite[approximately $15^{\prime\prime}$ at a distance of $414$ pc,][]{Goicoechea2016}. Nevertheless, previous infrared observations at 15$^{\prime\prime}$ resolution show extended C$^+$ emission along the Horsehead and IC 434 front \citep{Pabst2017,Bally2018}. Since C$^+$ observations are a direct tracer of the PDR edge, then some of the CO emission we detect toward the cloud edge probably arises from gas layers where not all carbon is in its molecular form but in either C or C$^+$.

The above is consistent with what we see at the edge of the Horsehead, beyond the PDR position ($\delta x \lesssim 5^{\prime\prime}$ in our figures). In that part of the cloud, the emission does not seem to be very affected by self-absorption effects, since the line profiles are approximately Gaussian. Still, if the CO $J=3-2$ line is thermalized, we would expect \TpeakCO{} values as high as $\sim 80-90$~K since the more exposed layers should present higher temperatures \citep{Pety2005}, on the order of $T_{\mathrm{kin}} \gtrsim 100$~K (see Fig.~\ref{fig:hh-excitation-models}). Since we do not recover such high temperatures at the cloud edge (see Fig.~\ref{fig:hh-Tpeak}), we conclude that $n_{\mathrm{H}} < n_{\mathrm{cr}}$. Indeed, towards the surface layers of the Horsehead, a lower CO column density is expected \citep{Pety2005}, and we found that the observed values of \TpeakCO{} $\approx 30-40$~K at the edge of the cloud are compatible with the expected temperatures for gas densities of about $n_{\mathrm{H}} \sim 10^{3}$~cm$^{-3}$ and $N(\mathrm{CO}) \sim 1 \times 10^{17}$~cm$^{-2}$ (red line, left panel in Fig.~\ref{fig:hh-excitation-models}). Both values are lower than those foreseen at the PDR position from C$^{18}$O observations \citep{Pety2005}, which supports the presence of a steep density gradient as previously suggested for the Horsehead edge \citep{Habart2005}. The approximation of \TpeakCO{} to $T_{\mathrm{kin}}$ is therefore no longer valid in this region.

Predictions from PDR models also suggest a similar scenario. We verified that stationary cloud models with constant gas density can reproduce the observed distance between the ionization and dissociation fronts in the Horsehead, but fail to reproduce the location of the C/CO transition zone. This suggests that the density is not constant but a steep density gradient is present at the edge of the PDR. Indeed, isobaric stationary models can reproduce the separation between the different transition zones with a representative thermal pressure of $P_\mathrm{th} = 6\times 10^6$ K cm$^{-3}$ for a radiation field of $G_0 = 100$ (see Fig.~\ref{fig:hh-CompatibleModels}), consistent with our estimates from the ALMA images and the presence of a density gradient \citep{Habart2005}. Other PDRs are also generally best modeled with a constant thermal pressure \citep{Marconi1998,Joblin2018}. In particular, highly UV-illuminated and denser PDRs are best modeled with an isobaric prescription but with larger thermal pressures \citep[$P_\mathrm{th} \sim10^8$ K cm$^{-3}$,][]{Joblin2018}. 

Stationary models are, however, a simplification of real PDRs which are most likely affected by dynamical processes induced by the UV radiation field \citep{Goicoechea2016,Joblin2018}. Actually, different star formation and stellar feedback numerical codes that simulate the expansion of an \HII{} region using hydrodynamics agree that strong density contrasts are present close to the IF \citep{Bisbas2015}, consistent with what we observe here thanks to the high angular resolution. Dynamical effects have also been proposed to explain the discrepancy between the modeled and observed H$_2$ line intensities in PDRs \citep{Habart2011}. Indeed, the intensities of the rotationally excited H$_2$ lines can be underestimated in low and moderately excited PDRs \citep{Habart2004,Habart2011}.

\subsection{Dynamical effects}

Far-ultraviolet radiation plays an important role in setting the thermal gas pressure in star-forming clouds at all spatial scales, which is illustrated by the apparent $G_0-P_{\mathrm{th}}$ correlation found in diverse environments, from local PDRs located at the rims of molecular clouds to entire star-forming galaxies \citep{Wolfire2022}. In the Horsehead case, we estimate the thermal pressure at the PDR position to be $P_\mathrm{th} = n_{\mathrm{PDR}}T_{\mathrm{PDR}} = \left(2.3-4.0\right)\times 10^6$ K cm$^{-3}$, which agrees with estimations toward other PDRs spanning different values of $G_0$ \citep{Joblin2018}. However, our $P_\mathrm{th}$ value is higher than what is expected in an \HII{} region and PDR interface that is at pressure equilibrium \citep{Seo2019}. In fact, by comparing with the estimated pressure in the IC 434 \HII{} region \citep{Bally2018}, we find $P_\mathrm{th,PDR} > P_{\mathrm{th,\mathrm{\HII{}}}}$. Based on these out of equilibrium values, and the compression factor that we estimated in Sect.~\ref{subsec:Density}, we conclude that dynamical effects are probably relevant in the Horsehead PDR, and that compression and photoevaporation may be at work, as previously suggested by observations of the ionized gas \citep{Bally2018} and dynamical PDR models \citep{Maillard2021}.

Moreover, according to our definition of the different transition zones (see Fig.~\ref{fig:hh-fronts}), there seems to be a correlation between the position of the fronts and the variation of the edge of the cloud (see right panel, Fig.~\ref{fig:hh-distances}). Interestingly, the above correlation has an apparent vertical oscillation with a certain periodicity. Indeed, by using a Lomb-Scargle periodogram, we detected a periodic signal with a period of approximately $23^{\prime\prime}$, equivalent to physical scales of about $0.05$~pc. A similar kind of oscillatory behavior has been witnessed before on the surface of the Orion molecular cloud, which is likely associated with a Kelvin-Helmholtz instability \citep{Berne2010}. Another possible explanation could be the presence of a thin shell instability, where the shell of an spherical expanding shock undergoes an oscillatory fragmentation due to dynamic and gravitational perturbations \citep{Vishniac1983}. Thus, our results could be an observational proof of dynamical effects at the edge of the Horsehead nebula.

Even though our results show that stationary models are compatible with the observations, the distance between the IF and the H/H$_2$ front measured with the observations presented here is only an upper limit, then any model predicting a smaller separation between these two fronts would also be compatible. In particular, PDR models that include dynamical effects have found that the size of the atomic layer can be significantly reduced compared to stationary models. The IF and H/H$_2$ transition can even merge in a single front \citep{Maillard2021}. Low excitation PDRs, like the Horsehead, have been found to be more affected. We therefore cannot discriminate between a stationary PDR model and a dynamical PDR with a reduced IF-H/H$_2$ distance (or even merged fronts). Consequently, we cannot rule out that dynamical effects are at play in the Horsehead. Higher angular resolution observations are needed to better estimate the distances between the H/H$_2$ and H$^+$/H transition fronts, and determine the importance of dynamical effects. Future observations with the Mid-InfraRed Instrument (MIRI) on board JWST of the H$_2$ rotational lines will also be key to asses the importance of dynamical effects in the Horsehead as well as in other PDRs \citep{Berne2022}. 

\section{Summary}\label{sec:Summary}
We have obtained the sharpest images of the molecular gas emission in the Horsehead nebula. The new ALMA images have revealed the fundamental, small scale, structure of the gas at the midly UV-illuminated rim of a molecular cloud, which is characterized by a web of bright filaments embedded in an extended fainter cloud emission. Under stellar UV irradiation conditions and dust content typical of Milky Way star-forming regions, the rims of molecular clouds such as the Horsehead nebula show steep density gradients and a very sharp transition between the molecular gas and the ionized gas. This emphasizes the role of CO as a powerful tracer of the bulk molecular gas physical conditions. Based on the measured distances between the different transition zones, our observations suggest that the atomic layer is very thin ($<650$~au) and the amount of CO-dark H$_2$ gas in the Horsehead PDR is very small ($A_{\mathrm{V}} = 0.006 - 0.26$ mag). The observed distances between the ionization and dissociation fronts, as well as between the ionization front and the C$^+$/C/CO transition zone, are well reproduced by isobaric stationary models with a UV radiation field between $G_0 = 50 - 186$, based on previous estimates, and thermal gas pressures of $P_{\mathrm{th, model}} = (3.7-9.2)\times10^{6}$ K cm$^{-3}$, which is consistent with our estimates from the physical conditions of the gas, $P_{\mathrm{th, obs}} = \left(2.3 - 4.0 \right) \times 10^6$~K~cm$^{-3}$. However, based on the pressure imbalance between the \HII{} region and the molecular cloud surface, we cannot rule out dynamical effects with the current observations. Future JWST observations of the intensity of the H$_2$ rotational lines will help to determine the importance of dynamical effects in PDRs. Additionally, taking into account the evidence of compression, possible ALMA observations of other molecular lines could be explored to determine the temperature and density in the compressed area, such as H$_2$CO which has quite strong lines that can be combined to get either the gas density or the gas temperature \cite[e.g.,][]{Guzman2011}.

\begin{acknowledgements}
The authors thank the referee for the constructive comments that improved the content of this work. We also thank the ALMA Data Analysts and Support Scientists from the North American ALMA Science Center (NAASC) for their support in data reduction. This paper makes use of the following ALMA data: ADS/JAO.ALMA\#2019.1.00558.S. ALMA is a partnership of ESO (representing its member states), NSF (USA) and NINS (Japan), together with NRC (Canada), MOST and ASIAA (Taiwan), and KASI (Republic of Korea), in cooperation with the Republic of Chile. The Joint ALMA Observatory is operated by ESO, AUI/NRAO and NAOJ. The National Radio Astronomy Observatory is a facility of the National Science Foundation operated under cooperative agreement by Associated Universities, Inc. C.H.-V. acknowledges support from the
National Agency for Research and Development (ANID) -- Scholarship Program through the Doctorado Nacional grant no. 2021-21212409. 
V.V.G. gratefully acknowledges support from FONDECYT Regular 1221352, ANID BASAL projects ACE210002 and FB210003, and ANID, -- Millennium Science Initiative Program -- NCN19\_171.
J.R.G. thanks the Spanish MICINN for funding support under grant PID2019-106110GB-I00.
V.M., J.P., F.L.P., M.G., E.B., E.R., P.G. acknowledge support by the Programme National “Physique et Chimie du Milieu Interstellaire” (PCMI) of CNRS/INSU with INC/INP, co-funded by CEA and CNES. 
\end{acknowledgements}

\bibliographystyle{aa} 
\bibliography{main}

\begin{appendix} 
\section{Channel maps}\label{App:ChannelMaps}

\begin{figure*}[ht]
\centering
\includegraphics[width=\linewidth]{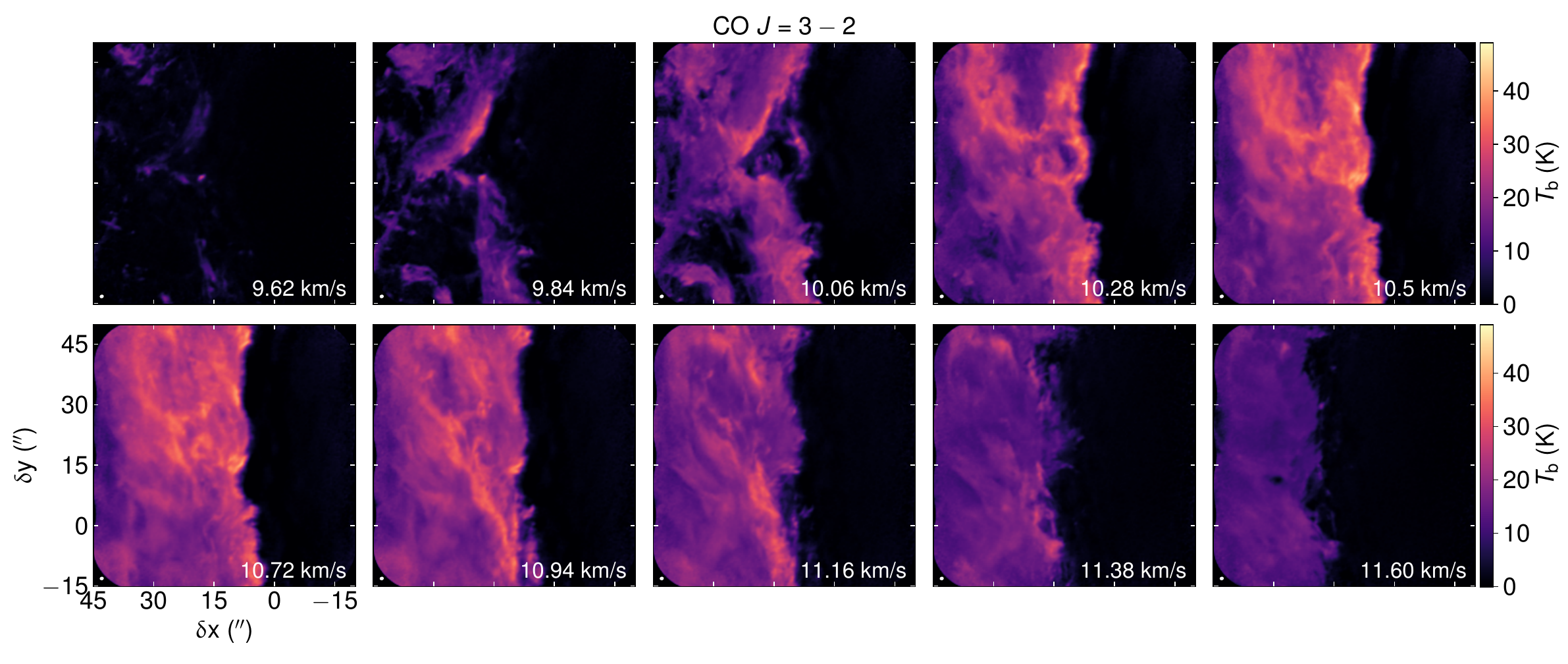}
\includegraphics[width=\linewidth]{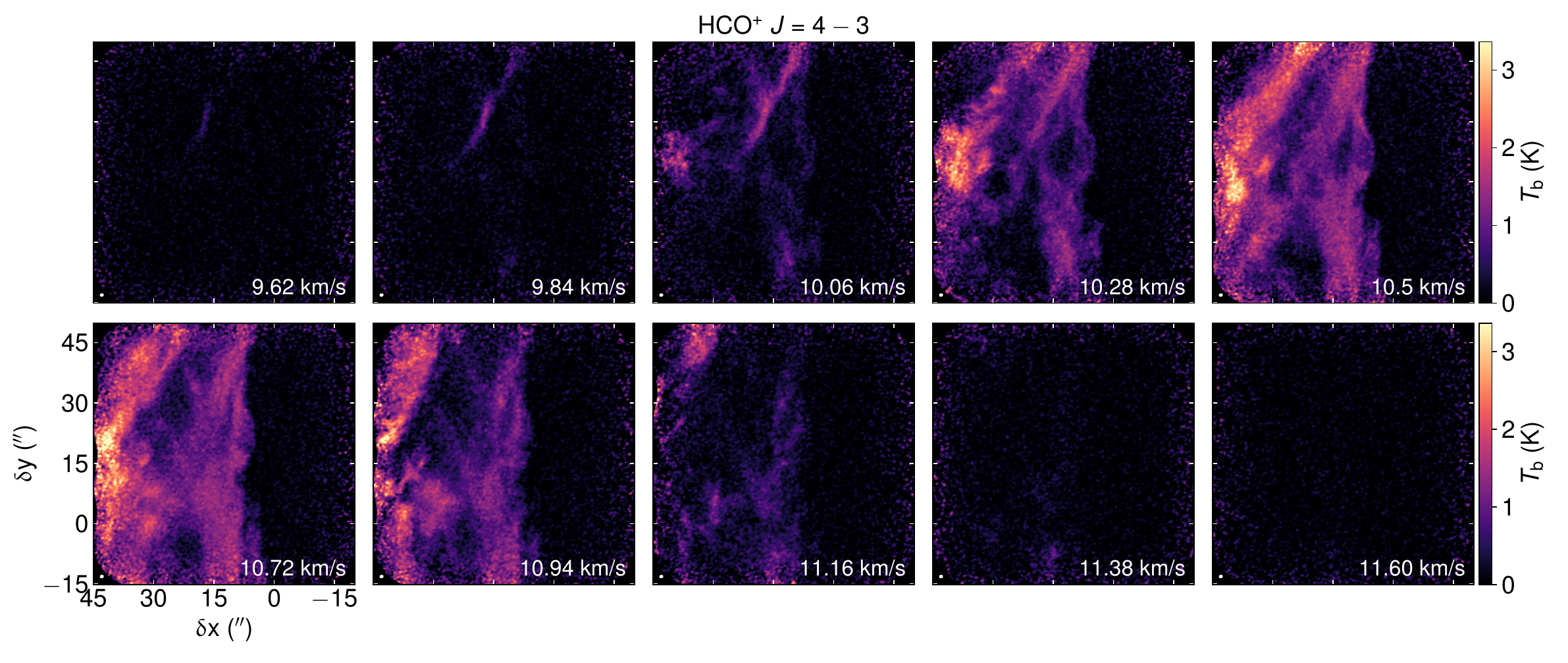}
\caption{Slices of the full ALMA data cubes, showing the CO $J=3-2$ (top) and HCO$^+$ $J=4-3$ (bottom) line emission at different velocities, ranging from $9.62$~km~s$^{-1}$ to $11.60$~km~s$^{-1}$ with a spectral resolution of $0.22$~km~s$^{-1}$. Data cubes were rotated $14^\circ$ counterclockwise to bring the illuminating star direction in the horizontal direction. The synthesized beam size and the kinematic local standard of rest (LSRK) velocities are shown in the bottom left and right corner of each panel, respectively.}
\label{fig:hh-channelmaps}
\end{figure*}

Several filamentary structures can be distinguished in the channel maps (see Fig.~\ref{fig:hh-channelmaps}). For example, a bright filament is clearly seen at the north of the image at 9.84~km~s$^{-1}$ in both CO and HCO$^+$. Another one is seen towards the south of the image at 10.94~km~s$^{-1}$, close to the PDR position, that is almost parallel to the ionization front. A few compact structures are also seen in CO, for example the one near the center of the image at low velocities (between $9.6 - 10.0$~km~s$^{-1}$). The edge of the Horsehead is seen at intermediate velocities, near the systemic velocity of about $10.5$ km s$^{-1}$. HCO$^+$ is brighter in the UV-shielded part of the cloud, close to the dense core position, but the line also shines at the UV-illuminated edge of the cloud, tracing a filament of dense gas that is parallel to the ionization front. 

\end{appendix}
\end{document}